\documentclass[12pt]{article}

\RequirePackage[OT1]{fontenc}
\usepackage{amsthm,amsmath,natbib}
\usepackage{array}
\usepackage{booktabs}
\usepackage{tabularx}
\usepackage{xcolor}
\RequirePackage[colorlinks,citecolor=blue,urlcolor=blue]{hyperref}
\usepackage{gensymb}

\def\bi{\begin{itemize}}
\def\ei{\end{itemize}}
\def\be{\begin{equation}}
\def\ee{\end{equation}}

\def\rm#1{\mathrm{#1}}

\usepackage[english]{babel}
\usepackage{amsfonts}
\usepackage{amssymb}
\usepackage{graphicx}
\usepackage{enumerate}
\usepackage{bm}

\def\ind{{\ {\buildrel \rm{ind}\over \sim}\ }}

\title{Point process-based modeling of multiple debris flow landslides using INLA: an application to the 2009 Messina disaster}

\author
{
	Luigi Lombardo$^{12*}$, Thomas Opitz$^3$, Raphael Huser$^1$\\
}
\date{\today}

\begin{document}
	\maketitle
 \begin{center}
{\large{\bf Abstract }} 
\end{center}
We develop a stochastic modeling approach based on spatial point processes of log-Gaussian Cox type for a collection of around $5000$ landslide events provoked by a precipitation trigger in Sicily, Italy. Through the embedding into a hierarchical Bayesian estimation framework, we can use the Integrated Nested Laplace Approximation methodology to make inference and obtain the posterior estimates. Several mapping units are useful to partition a given study area in landslide prediction studies. These units hierarchically subdivide the geographic space from the highest grid-based resolution 
to the stronger morphodynamic-oriented slope units. Here we integrate both mapping units into a single hierarchical model, by treating the landslide triggering locations as a random point pattern. This approach diverges fundamentally from the unanimously used presence-absence structure for areal units since we focus on modeling the expected landslide count jointly within the two mapping units. Predicting this landslide intensity provides more detailed and complete information as compared to the classically used susceptibility mapping approach based on relative probabilities. To illustrate the model's versatility, we compute absolute probability maps of landslide occurrences and check its predictive power over space. While the landslide community typically produces spatial predictive models for landslides only in the sense that covariates are spatially distributed, no actual spatial dependence has been explicitly integrated so far for landslide susceptibility. Our novel approach features a spatial latent effect defined at the slope unit level, allowing us to assess the spatial influence that remains unexplained by the covariates in the model. For rainfall-induced landslides in regions where the raingauge network is not sufficient to capture the spatial distribution of the triggering precipitation event, this  latent effect hence provides valuable imaging support on the unobserved rainfall pattern.


\footnotetext[1]{
\baselineskip=10pt Computer, Electrical and Mathematical Sciences and Engineering (CEMSE) Division, King Abdullah University of Science and Technology (KAUST), Thuwal 23955-6900, Saudi Arabia.}
\footnotetext[2]{
\baselineskip=10pt Physical Sciences and Engineering (PSE) Division, King Abdullah University of Science and Technology (KAUST), Thuwal 23955-6900, Saudi Arabia.}
\footnotetext[3]{
\baselineskip=10pt  INRA, UR546 Biostatistics and Spatial Processes, 228 Route de l'A\'erodrome, Avignon, France.}

\section{Introduction}

Landslide susceptibility maps are typically the result of spatial predictive models \citep{brenning2005spatial,chen2017landslide}. However, the spatial dimension in these models is often only carried through the observed predictors, which vary over space. No actual spatial dependence in a stochastic sense has so far been considered in the geomorphological literature despite the fact that the geostatistical community routinely investigates such spatial effects, which may account for unobserved predictors such as the spatially varying intensity of the precipitation trigger; for a comprehensive review of geostatistical models and methods, see, e.g., the books of \citet{Cressie:1993,Stein:1999,Wackernagel:2003,Diggle.Ribeiro:2007} and \citet{Cressie.Wikle:2011}. The principal aim of our contribution is to bridge the gap between the geomorphological and the geostatistical communities by accounting for latent spatial effects in landslide modeling, while predicting multiple debris flows scenarios. 

The model we advocate in this paper can essentially be represented as the Bayesian formulation of a Generalized Additive Model \citep[GAM;][]{Hastie.Tibshirani.1990,brenning2008statistical}. However, we propose several modifications to the current literature \citep[see, e.g.,][]{goetz2011integrating}. The primary difference resides in the probability distributions fitted to describe the landslide scenario. The landslide community unanimously pursues a binary presence-absence set-up corresponding to a Bernoulli probability distribution. Alternatively, we will here rely upon a Poisson probability distribution for event counts in small-area units. The Poisson distribution characterizes the random number of events contained in a given spatial unit, which extends the common binary situation. More precisely, we work with a spatial point process model defined over continuous space, which allows us to aggregate event counts and probabilities over any area of interest, independently of the initial spatial pixel discretization used for estimation. A major benefit for landslide modeling is the possibility to calculate the spatial landslide \emph{intensity} in addition to the \emph{susceptibility}. While the latter represents the relative (and potentially rescaled) probability of observing at least one landslide in a given mapping unit, the former indicates the expected number of landslides in such a unit \citep{erener2012landslide}; the intensity therefore contains more information than the susceptibility for landslide risk assessment.

In the geomorphological literature, there is often a misunderstanding that the notion of susceptibility actually corresponds to the \emph{exact} probability of observing a landslide in a given mapping unit. However, susceptibility maps are often calculated from estimated logistic regression models, which are typically fitted to an artificially created, balanced \citep{lombardo2014test} or unbalanced \citep{heckmann2014sample}, dataset of landslide presences and absences. Therefore, instead of characterizing the ``true'' landslide probability, the susceptibility rather describes a sort of relative likelihood of unstable versus stable terrain conditions. \citet{petschko2014assessing} recognized this misconception and clearly stated the fundamental difference between ``true'' probability and susceptibility; the authors first compute the odds based on a balanced dataset and then correct them to obtain the ``true'' probability odds. Using our point process approach, ``true'' probabilities can be naturally and directly calculated for any area of interest.

Another key feature of our proposed modeling approach is to account for spatially correlated unobserved factors, which affect a given landslide scenario. We assume that available (i.e., observed) covariates do not explain the whole spatial variability of landslide occurrence. Part of the unexplained component may be captured via a latent spatial effect. Here, we try to characterize the observed and latent effects in a Bayesian modeling framework with a log-Gaussian Cox point process using Integrated Nested Laplace Approximation \citep[INLA,][]{Rue.Martino.Chopin.2009,Illian.al.2012,Rue.al.2016} for fast and accurate estimation. 


Compared to the current geomorphological literature, we also propose a different take on ordinal predictors, often discretized into a number of categories treated as independent. The community constantly uses estimation techniques that neglect the ordered structure of the categories, whose integration may considerably improve estimation and interpretation through the realistic assumption that neighboring categories tend to have similar estimated coefficients, as compared to distant ones. The traditional assumption of independent categorical classes is surely valid when using geology or land use; however, it may represent a gross simplification and loss of information for covariates such as the aspect, or the slope, which are often transformed into corresponding categorical variables. In this contribution, we account for the internal dependence between adjacent categorical classes.

Our novelties fit into the big picture of methodological developments for more precise and informative modeling of high-dimensional geomorphological datasets, bridging state of the art in geostatistical Bayesian inference and geomorphology. However,~we also present novel and purely geomorphological considerations for the studied dataset. A number of papers have been published on the disaster that occurred in Messina, Italy, in 2009, focusing only on the catchments that sustained most of the damages \citep{cama2016exploring,lombardo2015binary}, and only few cases take into account the basins at the margin of the storm \citep{lombardo2016a,zini2015rusle}. In this paper, we develop a statistical model for the entire region affected by the landslides triggered on October 1, 2009, as a result of heavy rainfall, and we study the effects of both the predisposing factors and the precipitation trigger event. We offer a detailed discussion of this natural disaster, while highlighting how advanced statistical models and methods may be more broadly applied to assess the risk of landslide in other regions and contexts.

The following Section~\ref{sec:data} presents available data and preprocessing steps. Hierarchical statistical  modeling with log-Gaussian Cox processes is developed in Section~\ref{sec:modeling}, and we provide   the description of the specific models that we propose for our data. We discuss important statistical details of estimation results  in Section~\ref{sec:estimation}. Owing to a number of innovating model features as compared to state-of-the-art approaches in landslide modeling, we provide a profound geomorphological interpretation of these statistical results in a separate Section~\ref{sec:interpretation} before concluding the paper in Section~\ref{sec:conclusion}.

\section{Landslide inventory and geomorphological dataset}
\label{sec:data}
\subsection{Study region}
\begin{figure}[t!]
	\centering
	\includegraphics[width=\linewidth]{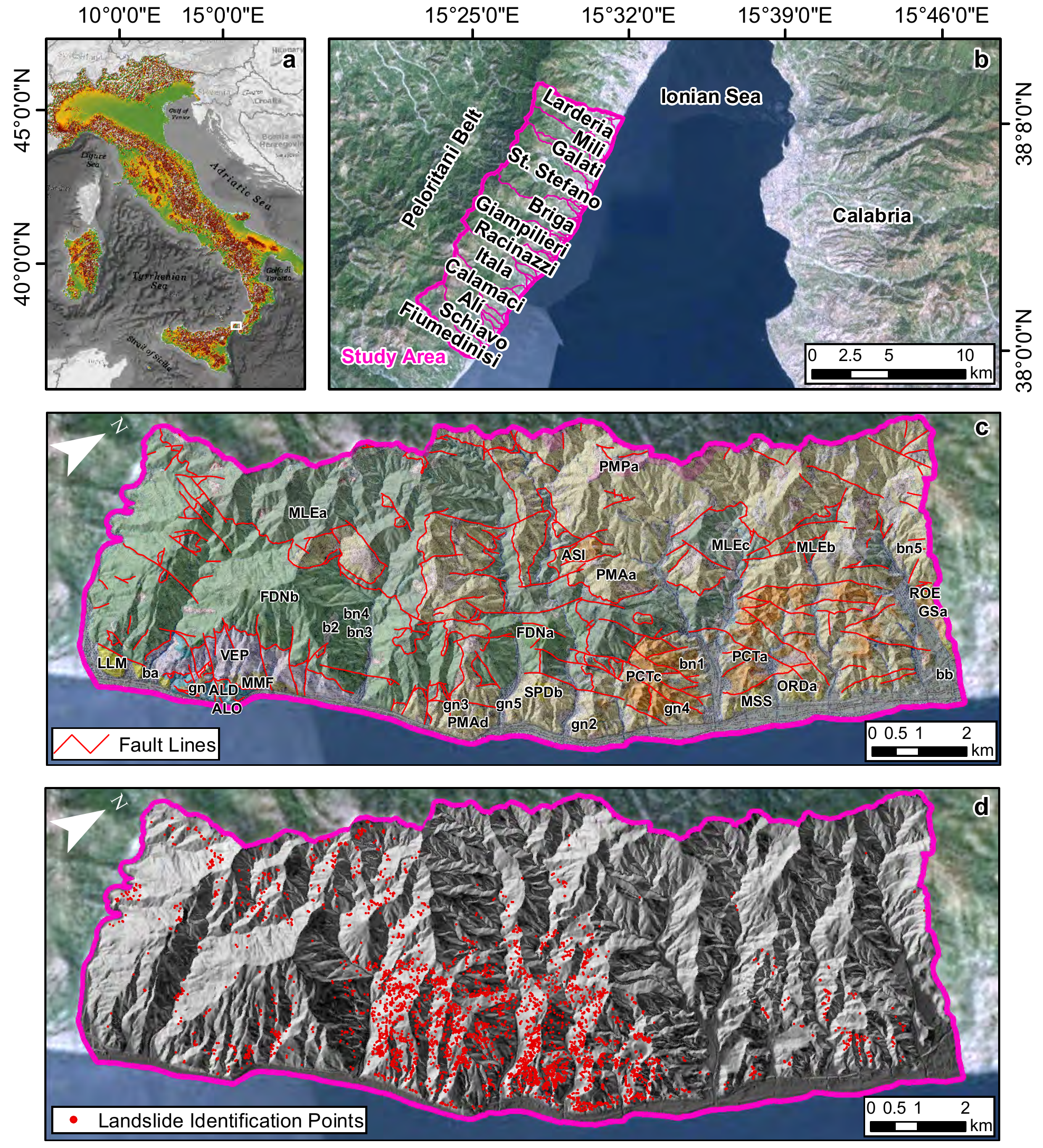}
	\caption{Italy (a); Study area (b); Outcropping Lithologies and tectonic lineaments (c); Landslide inventory (LIPs) on October 1, 2009 (d).}
	\label{fig:map}
\end{figure}

Our region of study encompasses twelve catchments spanning from the southernmost Fiumedinisi to the northernmost Larderia (see Figure~\ref{fig:map}). These catchments correspond to the geographic entities that suffered from landslide activations on October 1, 2009. In particular, $68.3\%$ of the mass movements are actually concentrated in the epicentral sector corresponding to the catchments of Itala, Racinazzi, Giampilieri and Briga, but because our aim is to model the spatial effect due to the precipitation trigger, we include all the basins up to the margins of the storm. The catchments to the south, namely Fiumedinisi, Schiavo, Al\'i and Calamaci, account for $26.3\%$ of total landslides whereas Santo Stefano, Galati, Mili and Larderia to the north suffered the remaining $5.4\%$. Minor intermediate catchments are aggregated to the adjacent major ones in these estimations. These percentages may already give indications of the spatial evolution of the main storm, but it will be crucial to correctly disentangle geomorphological effects susceptible to present strong spatial heterogeneity (morphometry, landforms, lithology, land-use) from the effect of the intensity of the trigger. The analysis of \citet{aronica2012flash} has already revealed that an initial cloudburst to the south quickly migrated to the center of the study area where the convective system released the majority of the total discharge. 

Measurements at the weather station of Briga close to the epicenter report $250$mm of rain on the day of the disaster, in addition to two smaller precipitation events ($190$mm and $75$mm in a single day) that took place one and two weeks before the main event, respectively. In this situation, the weathered mantle draping over a medium to high metamorphic bedrock became saturated and brought to the brink of instability, owing to this very intense climatic stress, unprecented over the past 30 years  \citep{cama2015predicting}. The bedrock primarily consists of paragneiss, gneiss, micaschists and phyllites, which have been weakened by several tectonic cycles piling up the lower groups into numerous duplexes \citep{giunta1996nuove} and slicing the uppermost groups with synorogenic normal faults \citep{somma2005syn}. 
Consideration should also be given to the catchment morphology  since the distance of around $7$km from the coastline to the highest ridge is associated with a $1$km variation in elevation, making catchments short and steep. The steepness reaches peak values of $86\degree$ with a mean of $29\degree$ and standard deviation of $13\degree$. These values already indicate a general morphology prone to fail since slopes steeper than $20\degree$ are often reported to be the triggering threshold for debris flows under unfavorable conditions \citep[e.g.,][]{imaizumi2006hydrogeomorphic}.  

\subsection{Landslide inventory}
Several remote sensing scenes have been analyzed to create a full inventory \citep{malamud2004landslide} of the 2009 Messina event. The initial interpretation relied on post-event orthophotos at $0.17$m resolution provided by the Italian National Civil Protection (PCN). Landslide signatures were refined on the basis of Kompsat-2 scenes at $4$m resolution together with information from Google Earth, ESRI and Bing Basemaps. A comparison with pre-event orthophotos provided by the Territory and Environment Department (ARTA) of the Sicilian Regional Council at $0.25$m resolution allowed for isolating the slope responses belonging to the $2009$ event. As a result, detailed polygon-shaped landslide scars were identified in the epicentral area, along with an additional landslide database for the marginal sectors consisting of the highest points along the landslide crowns. 
In order to generate a global homogeneous landslide inventory, we first resampled each landslide polygon from the epicentral area into points at a $2$m interdistance. 
Second, we extracted the planar coordinates of the highest point in each polygon, producing so-called Landslide Identification Points (LIPs); see, e.g., \citet{lombardo2014test} for full details on this procedure. Our final inventory contains $4879$ LIPs in the entire area, which are represented in Figure~\ref{fig:map}(d).

\subsection{Covariates}
We selected and preprocessed thirteen covariates to support subsequent statistical analyses. Nine of them are computed from a pre-event Digital Elevation Model (DEM) generated during a Light Detection and Ranging (LIDAR) survey in 2008\footnote{\url{http://www.sitr.regione.sicilia.it/geoportale/it/metadata/details/502}}. The Normalized Difference Vegetation Index (NDVI) is computed from the ASTER scene acquired on May 10, 2009. In addition, we include the distance to tectonic fault lines shown in Figure~\ref{fig:map}(c), the outcropping 
lithology (see Figure~\ref{fig:map}(c)) from the local $1$:$25000$ scale geological map\footnote{\url{http://www.isprambiente.gov.it/Media/carg/601_MESSINA_REGGIO/Foglio.html}} and the land use\footnote{\url{http://www.eea.europa.eu/data-and-maps/data/clc-2006-raster}}. The full list of covariates is summarized in Table \ref{table1}. Predictors represented as categorical covariates in the geomorphologic literature have always been  used without taking into account the ordinal or neighborhood relationships between categories, i.e., by considering categories as completely independent. This simplification may be reasonable for the lithology or land use, but other common predictors may strongly deviate from the initial hypothesis of independence. Therefore,  important information can be borrowed from neighboring categories during parameter estimation, and in doing so, a higher number of categories can be used to refine estimated effects. Alternatively, it would also be possible to consider a continuous function represented through B-spline basis functions for continuous ordinal covariates.  An important example covariate is the aspect (Asp), which denotes a planar angle in $[0, 360)$ reflecting the orientation of the slope with respect to the North; it is cyclic (as $\mbox{Asp}=0^\circ$ is equivalent to $\mbox{Asp}=360^\circ$), and neighboring classes are linked to each other. Furthermore, any reclassification of a continuous ordinal covariate will produce a new categorical one where each level is dependent on the previous and subsequent ones. Notice that such variable transformations are common in the statistical literature, where a single continuous variable is sliced into several classes to assess nonlinear effects while keeping the inter-class dependence, for instance through a Bayesian prior specification with inter-class dependence.  In this work we will check both scenarios of linearity (through fixed effects) and nonlinearity (through random effects) for each continuous covariate in Table~\ref{table1}. 

\begin{table}[t!]
	\caption{List of geomorphological covariates used in our statistical analysis, their original types, their acronyms, and their units.}
	\vspace{5pt}
\label{table1}
\centering
\begin{tabular}{|l|c|c|c|} 
 \hline
 \textbf{Covariate} & \textbf{Original Type} & \textbf{Acronym} & \textbf{Unit} \\ 
 \hline
 Aspect & Cyclic Categorical & Asp & degree ($^\circ$)\\ 
 Distance to Faults & Continuous & Dist2F & meter (m)\\
 Elevation & Continuous & Elev & meter (m)\\ 
 Landform Classification & Categorical & LandC & \textit{unitless}\\
 Land Use & Categorical & Use & \textit{unitless}\\
 Normalized Difference Vegetation Index & Continuous & NDVI & \textit{unitless}\\
 Outcropping Lithology & Categorical & Litho & \textit{unitless}\\
 Planar Curvature & Continuous & Plc & m$^{-1}$\\
 Profile Curvature & Continuous & Prc & m$^{-1}$\\
 Relative Slope Position & Continuous & RSP & \textit{unitless}\\
 Slope & Continuous & Slo & degree ($^\circ$)\\
 Stream Power Index & Continuous & SPI & \textit{unitless}\\
 Topographic Wetness Index & Continuous & TWI & \textit{unitless}\\
 \hline
\end{tabular}
\end{table}

\subsection{Mapping units}

\begin{figure}[t!]
	\centering
	\includegraphics[width=\linewidth]{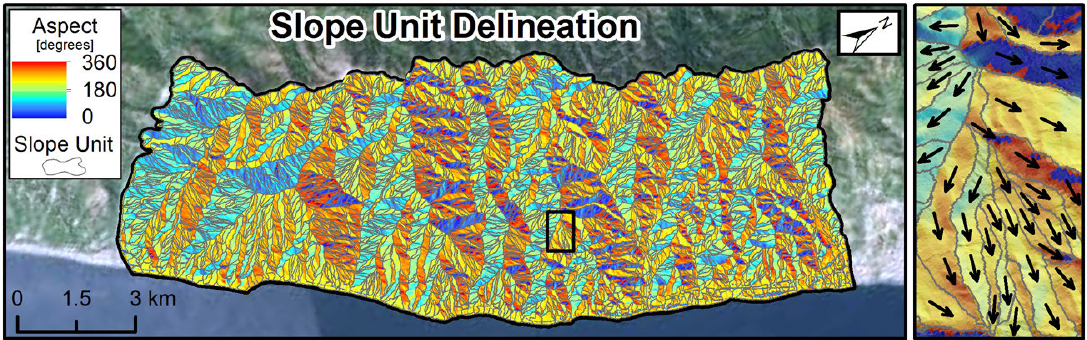}
	\caption{Slope Units partition underlain by Aspect. The right panel shows a zoom where arrows indicate the interpreted path of a potential debris flow for each slope unit. This highlights the independence of the slope response in neighboring slope units.}
	\label{fig:slopeunits}
\end{figure}

We focus on two types of mapping units: (i) a high-resolution regular grid comprising $449250$ squared pixels of area $225$m$^2$ covering the entire study region; and (ii) slope units \citep{carrara1995gis}. Covariates are expressed at the fine spatial grid resolution; if a covariate was initially available at a higher resolution, we resample it at the $15$m cell-size by computing its mean value over the coarser pixel. This reduces computational costs, while keeping the mapping units to a reasonably small size, shown to produce good results for our specific study region; see \citet{arnone2016effect,cama2017improving} and \citet{lombardo2016b}. The partitioning into slope units (see Figure \ref{fig:slopeunits}) was achieved using the \texttt{r.slopeunit} software\footnote{\url{http://geomorphology.irpi.cnr.it/tools/slope-units}}, and parametrized as in \cite{alvioli2016automatic} and \cite{rossi2016land} for the catchment of Giampilieri. As a result, we generate a total number of 3848 slope units.

In our subsequent statistical analysis, the latent spatial effect is defined at the slope unit level. Slope units therefore play a key role in the present paper. Defining the latent spatial effect over the high-resolution grid would not only be extremely computationally demanding, but the strong dependence between neighboring pixels would also induce numerical instabilities. Slope units are also attractive from an interpretation perspective, because they form geomorphologically independent spatial entities that provide a homogeneous response of a given slope when a landslide occurs; see the illustration on the right panel of Figure~\ref{fig:slopeunits}.

\section{Hierarchical Bayesian modeling of point patterns}
\label{sec:modeling}
\subsection{Spatial point processes for random point patterns}
The response variable to model and predict consists of the occurrence positions of the landslide events (i.e., the LIPs shown in Figure~\ref{fig:map}(d)), considered as random because they are unknown before the event takes place. The natural approach for  representing a collection of random points over continuous space is through spatial point processes whose realizations are point patterns. We here discuss their appealing properties for predicting event probabilities and intensities over arbitrary spatial supports. Event positions are random but they may be more or less dense in certain regions of space, and they may exhibit clustering or repulsion properties at small scales. In the landslide context, the modeling of clustering structures due to unobserved effects (such as the precipitation trigger in our case) is of principal interest.  

The main characteristic of a point process is its intensity function $\lambda(s)\geq 0$: up to a constant factor, the intensity $\lambda(s)$ may be interpreted as the expected number of points falling into an infinitesimal region around the location $s$. More precisely, the count of events occurring in an area $A$ is a non-negative random variable $N(A)\geq0$, whose expectation is given by the integral of the intensity function over $A$, i.e., $E(A)=\int_A \lambda(s)\, \mathrm{d}s$. 

The fundamental point process model is the Poisson point process, which is characterized by two probabilistic properties: (i) the number of events $N(A)$ occurring in a bounded area $A$ follows the Poisson distribution, i.e., $\mathrm{pr}\{N(A)=k\}=\exp\{-E(A)\} E(A)^k/k!$, $k=0,1,\ldots$; (ii) if $A_1$ and $A_2$ denote two disjoint areas of space, then $N(A_1)$ and $N(A_2)$ are independent. In other words, the events are randomly scattered over space, yet according to the Poisson distribution defined in terms of a \emph{deterministic} intensity $\lambda(s)$ that may vary spatially. In particular, the probability of absence of any event in $A$ is given by $\mathrm{pr}\{N(A)=0\}=\exp\{-E(A)\}=\exp\{-\int_A \lambda(s)\, \mathrm{d}s\}$. The simplest case consists of homogeneous Poisson processes, which have a constant intensity $\lambda(s)\equiv \lambda$: the expected number of events in an area $A$ is $E(A)=\lambda|A|$, and the probability density of the occurrence position for a point picked at random from overall $N(A)$ points is the uniform density $1/|A|$ over the area $A$. 

This probabilistic framework based on point processes is richer than the dichotomous presence-absence setting for specific areal units, fixed a priori, either based on a fine-scale regular grid (at pixel resolution) or larger-scale administrative or geological areal units (such as slope units or catchments). In classical geomorphology-driven landslide modeling, estimation and prediction is tailored to the concept of susceptibility by contrasting areal units touched by landslides with a random selection of the same number of the untouched units. Through its capability to provide probabilities of the type $\mathrm{pr}\{N(A)=0\}$ for any area $A$, our modeling paradigm may be used to derive susceptibility maps in the traditional sense, but it also gives valuable additional information on the point intensity over space and probabilities of the type $\mathrm{pr}\{N(A)=k\}$ or  $\mathrm{pr}\{N(A) > k\}$ for $k=0,1,\ldots$, and any areal unit $A$. 

\subsection{Log-Gaussian Cox processes for random effect modeling}
A spatial Cox point process is essentially a Poisson point process with a \emph{random} intensity function, denoted $\Lambda(s)\geq 0$, $s\in S$. In other words, conditional on $\Lambda(s)=\lambda(s)$, we obtain a Poisson point process with intensity $\lambda(s)$. Through their doubly stochastic construction, Cox point processes are natural models for capturing clustering of points, i.e., the fact that even after taking into account the effect of observed covariates in the Poisson intensity, there remain areas with relatively higher or lower point intensity. In such cases, the point pattern shows clustering structures that cannot be fully explained by the available covariates; however, this may be modeled by assuming that the intensity function is random and contains a latent spatial effect that governs the remaining variation of the intensity over space. 

The Gaussian process is the workhorse of spatial statistics, while the log-Gaussian Cox process has risen as its counterpart for modeling point patterns. In this model, the Poisson intensity $\Lambda(s)$ is assumed to be a log-Gaussian process, allowing for the inclusion of fixed and random effects. More precisely, the log-intensity $\log\{\Lambda(s)\}$ is described as a Gaussian random field, in which fixed covariate effects and random effects are embedded through an additive structure of the form
\begin{equation}\label{eq:gauss}
\log\{\Lambda(s)\} = \beta_0+\sum_{j=1}^J \beta_j z_j(s) + \sum_{k=1}^K W_{\bm z_k}(s),
\end{equation}
where $\beta_0$ is the intercept, $z_j$ are fixed (linear) covariate effects with coefficient $\beta_j$, $j=1,\ldots,J$, and $W_{\bm z_k}$ encodes additional (nonlinear) random effects according to some functional form defined with respect to a covariate set $\bm z_k$, $k=1,\ldots,K$. 
Notice that we require the vector $\{W_{\bm z_k}(s_1),\ldots,W_{\bm z_k}(s_d)\}^T$ to be multivariate Gaussian for any set of sites $\{s_1,\ldots,s_d\}\subset S$. 

Specifically, for $k=1$, we model the spatial random effect $W_{\bm z_1}$ in \eqref{eq:gauss} as a conditional autoregressive \citep[CAR;][]{Besag.1975,Rue.Held.2005} Gaussian process defined over the study region at the slope unit level. Hence, $\bm z_1$ denotes here the collection of slope units, $W_{\bm z_1}$ represents a random vector (one variable per slope unit) described by a multivariate Gaussian distribution, and $W_{\bm z_1}(s)$ extracts the Gaussian variable associated to the slope unit containing the location $s$. The CAR model further specifies the Gaussian mean and covariance structures. It is intrinsically defined through the conditional relation of the Gaussian variable $W_\ell$ associated to the $\ell$th slope unit with the variables $W_{m}$ of neighboring slope units (i.e., sharing a common border with the $\ell$th slope unit):
\begin{equation}\label{eq:spateff}
W_\ell\mid W_{\bm z_1}\sim \mathcal{N}\left(\frac{1}{n_\ell}\sum_{m\in\mathrm{NB}(\ell)} W_m, \frac{1}{n_\ell\tau_1}\right),
\end{equation}
where $\mathrm{NB}(\ell)$ is the set of size $n_\ell=|\mathrm{NB}(\ell)|$ comprising the indices of neighboring slope units. Thus, the spatial effect for the $\ell$th slope unit is the mean of its neighboring slope units, with some additional noise of variance $1/(n_{\ell}\tau_1)$. The precision hyperparameter $\tau_1>0$ is crucial as it determines the spatial dependence strength, i.e., whether the random effects of neighboring slope units are strongly or weakly correlated, and we estimate it from the data.

The other random effects for $k=2,\ldots,K$ in \eqref{eq:gauss} correspond to nonlinear covariate effects. While a nonlinear model is clearly necessary for the Aspect (treated as a cyclic covariate), other continuous covariates such as the Slope or Distance to Faults may be considered as linear or nonlinear. By including additional nonlinear random effects for these covariates, our goal is to check whether or not the nonlinear component significantly improves the model. We here propose to discretize the continuous covariates into a sufficiently large number of equidistant bins such that $\bm z_k$ denotes a list of $L_k$ bins, $W_{\bm z_k}=(W_{\bm z_k,1},\ldots,W_{\bm z_k,L_k})^T$ is a Gaussian random vector of size $L_k$ defined on the bins, and $W_{\bm z_k}(s)$ extracts the random variable associated to the bin corresponding to the specific covariate value observed at location $s$. We further assume that $W_{\bm z_k}$ has first-order random walk structure, i.e., 
\begin{equation*}\label{eq:rw}
W_{\bm z_k,\ell}-W_{\bm z_k,\ell-1}\sim \mathcal{N}(0,1/\tau_k), \qquad \ell=2,3,\ldots,L_k.
\end{equation*}
Notice that we impose the condition $W_{\bm z_k,L_k}=W_{\bm z_k,1}$ for cyclic variables such as the Aspect. To ensure identifiability of the Gaussian variables in non-cyclic cases, we use the sum-to-zero constraint $\sum_{\ell=1}^{L_k} W_{\bm z_k,\ell}=0$. The hyperparameter $\tau_k>0$ determines the strength of dependence among neighboring covariate classes.

Closed-form expressions of the likelihood function  are not available for this model since it is not possible to integrate out the latent  log-Gaussian random effects in its density expression. Therefore, the use of Bayesian simulation-based techniques, either based on Markov chain Monte Carlo (MCMC) methods, or on astutely designed analytical approximations of the high-dimensional integrals involved, such as the approach of Integrated Nested Laplace Approximation \citep[INLA, ][]{Rue.Martino.Chopin.2009,Illian.al.2012,Rue.al.2016,Opitz.2017}, is common for estimating such complex models. 
In this work, we accurately estimate model parameters and predictive distributions by taking advantage of INLA (implemented in the R package \texttt{R-INLA})
, which bypasses the intricate updating schemes of simulation-based MCMC methods for high-dimensional and complex hierarchical models with non-gaussian responses. Appendix~\ref{sec:INLA} provides further details on Bayesian inference based on INLA and some guidance on the choice of prior distributions; for reproducibility purposes, we also make the \texttt{R} code available on GitHub; see \url{https://github.com/ThomasOpitz/popland}. 

\subsection{Poisson regression formulation}
We now describe a slightly modified formulation of the above point process model based on Poisson regression. It represents the standard inference approach for log-Gaussian Cox processes, which we use here, and it also provides an alternative interpretation of the model. 

Suppose that space is discretized into a grid of $n_{\mathrm{grid}}$ pixels of area $C$. Denoting by $N_i$ the number of landslides triggered in the $i$th pixel, $i=1,\ldots,n_{\mathrm{grid}}$, we can formulate a Poisson regression model conditionally on the intensity function $\Lambda(s)$ as follows:
\begin{equation}\label{eq:ppdiscr}
N_i\mid \Lambda(s) \ind \mathrm{Poisson}(C\Lambda(s_i)),\qquad i=1,\ldots, n_{\mathrm{grid}}.
\end{equation}
In comparison to the original point process formulation, we here assume that the Poisson intensity function $\Lambda(s)$ is constant within each pixel, which represents a negligible approximation if the pixel size is small. In the following, we choose the pixel according to the discretization of the environmental covariates, i.e., $C=(\mbox{15m})^2=225$m$^2$. Covariate influence can be estimated by embedding \eqref{eq:ppdiscr} into the framework of a generalized additive regression model. The multiplicative constant $C$ appears as an offset $\log(C)$ in the intercept of the regression. Using the approximation \eqref{eq:ppdiscr} remains possible with a random intensity
. Since the linear predictor containing fixed and random effects may be expressed as $X_i=\log\{\Lambda(s_i)\}=\beta_0+\ldots $ for pixel $i$ as in \eqref{eq:gauss}, 
we are using the canonical log-link function of classical Poisson regression. 

With this model, the probability $p_i$ of observing at least one event within the pixel $i$ is
\begin{equation}\label{eq:pi}
p_i=1-\exp\{-C\Lambda(s_i)\}=1-\exp\{-C\exp(X_i)\}, \quad i=1,\ldots, n_{\mathrm{grid}}.
\end{equation}
For relatively small probability values $p$, we may use the well-known approximation $\exp(p)\approx 1+p$, whose application to \eqref{eq:pi} yields  $p_i\approx C\exp(X_i)=C\Lambda(s_i)$. 
Notice that for logistic binary regression models traditionally used for presence-absence data, we have the following relationship between $p_i$ and the linear predictor $X_i$: 
\begin{equation}
p_i=  \frac{C\exp(X_i)}{1+C\exp(X_i)}, \quad i=1,\ldots, n_{\mathrm{grid}}.
\end{equation}
For small $p_i$, this also corresponds to the approximation $p_i\approx C\exp(X_i)$. Thus, the link between the linear predictor $X_i$ and the predicted probabilities is approximately the same in both modeling approaches (traditional one and ours) as long as the probabilities are small, which is usually the case when considering landslide events at a high pixel resolution, and provided the logistic regression model is fitted to a full (unbalanced) presence-absence dataset. 
However, we stress that our discretized Poisson model (and any more general point process model) provides important additional information by predicting also the probabilities for the exact number of events (i.e., for $k=0,1,2,\ldots$), and it allows coherent aggregation of intensities and probabilities over any areal support.   Moreover, by using count data and not only binary presence-absence information, the estimation algorithm also uses more precise information about the spatial event distribution, and one can expect a reduced estimation uncertainty, even when the ultimate goal is only to estimate presence-absence probabilities. 

For any areal unit $A$, a simple way of obtaining the fitted intensity $\hat\lambda(A)$ (i.e., the predicted number of events in $A$, which is an estimate of $C\int_A \Lambda(s)\,\mathrm{d}s$) and the fitted probability $\hat p(A)$ of observing at least one event in $A$ proceeds as follows. Aggregating the fitted intensities $\hat\lambda_i$ over all pixels included in $A$ yields the fitted intensity for $A$, i.e.,
\begin{equation}\label{eq:aggr}
\hat \lambda(A)=\sum_{s_i\in A} \hat\lambda_i.
\end{equation}
The event probability for $A$ can be estimated as $\hat p(A)=1-\exp\{-\hat \lambda(A)\}$. 

\subsection{Structure of fitted models}
We consider four models (denoted Mod1, Mod2, Mod2b and Mod3) of varying complexity, which incorporate covariates in a linear and possibly nonlinear fashion, and include or not a latent spatial effect. The Aspect is treated as a nonlinear covariate in every model, assuming a cyclic random walk structure over $16$ equidistant angle classes. To ensure identifiability of fixed effect coefficients for the non-ordinal categorical covariates related to lithology, land use and landforms, we impose sum-to-zero constraints on the corresponding regression coefficients.  We fix a precision parameter of $2$  (i.e., standard deviation $1/\sqrt{2}\approx0.7$) in the centered Gaussian prior for the fixed effects given by continuous covariates, except for the intercept with prior mean $-2$ and precision $1$. For the categorical fixed effects (lithology, land use, landforms) with more than $40$ categories overall, we use stronger priors with precision $100$ (i.e., standard deviation $0.1$). 

For the continuous covariates except Aspect, we allow for nonlinearities in some of the models by including both the fixed linear effect and a first-order random walk used to capture potential nonlinearities; if the latter is not significantly different from zero, then the simpler linear model is deemed to be more appropriate. In these random walk models, we slice the range of each continuous covariate into $20$ equidistant classes, assuming that the coefficients of neighboring classes are correlated. We fix the prior precision $\tau_k$  of the jump sizes between two neighboring classes to $25$ (i.e., standard deviation $0.2$), although it remains possible to obtain much larger jumps in the estimated values if the data provide strong evidence for it.  Decomposing the covariate effect into a linear and nonlinear part allows us to check separately for the presence of some monotonous effect of the covariate on the landslide intensity (linear part) and of more complex residual nonlinearities (random walk part). 

Since including such nonlinearities into the model makes its estimation more challenging, we proceed constructively as follows to build our four models: 
\begin{enumerate}
\item[i)] Mod1 has only linear effects, except for Aspect; 
\item[ii)] Mod2 has a nonlinear structure for all continuous covariates; 
\item[iii)] Mod2b has a nonlinear structure in some of the covariates, based on the inspection of the significance of the nonlinear components in Mod2; 
\item[iv)] Mod3 is partially nonlinear as Mod2b, and additionally includes a latent spatial effect defined through \eqref{eq:spateff}. 
\end{enumerate}

\begin{table}[t!]
\caption{Summary of fitted models.}
\vspace{5pt}
\label{table2}
\centering
\begin{tabular}{|c|c|c|} 
 \hline
 \textbf{Model ID} & \textbf{Covariate Use} & \textbf{Latent     Spatial Effect} \\ 
 \hline
 Mod1 & All linear except for Aspect & Absent \\ 
 Mod2 & All nonlinear & Absent \\
 Mod2b & Non-linear subset based on Mod2 & Absent \\
 Mod3 & Non-linear subset based on Mod2 & Present \\
 \hline
 \end{tabular}
 \end{table}

These four models are summarized in Table \ref{table2}. The subset of covariates with nonlinear effects selected for Mod2b and Mod3 is:  Elevation, Slope, Distance to Faults (and Aspect). Mod1, Mod2 and Mod2b are similar in spirit to the modeling paradigm routinely implemented in the current geomorphology literature, in which the spatial dimension enters into the model only through the observed covariates, whereas Mod3 is a significant step forward. 

We have endowed our models with sensible  choices of prior parameters, making them informative enough to avoid instabilities in the INLA-based estimation procedure. The only hyperparameter to be estimated is the precision of the spatial effect governing the strength of spatial dependence between slope units in Mod3. 
All continuous covariates (except for Aspect) have been rescaled prior to the analyses by subtracting their empirical mean value and dividing by their empirical standard deviation, which simplifies the comparison of fitted effects since they are expressed on the same unit-less scale.

\section{Estimation results}  \label{sec:estimation}

\subsection{Covariate effects}
\begin{figure}[t!]
	\centering
	\includegraphics[width=0.99\linewidth]{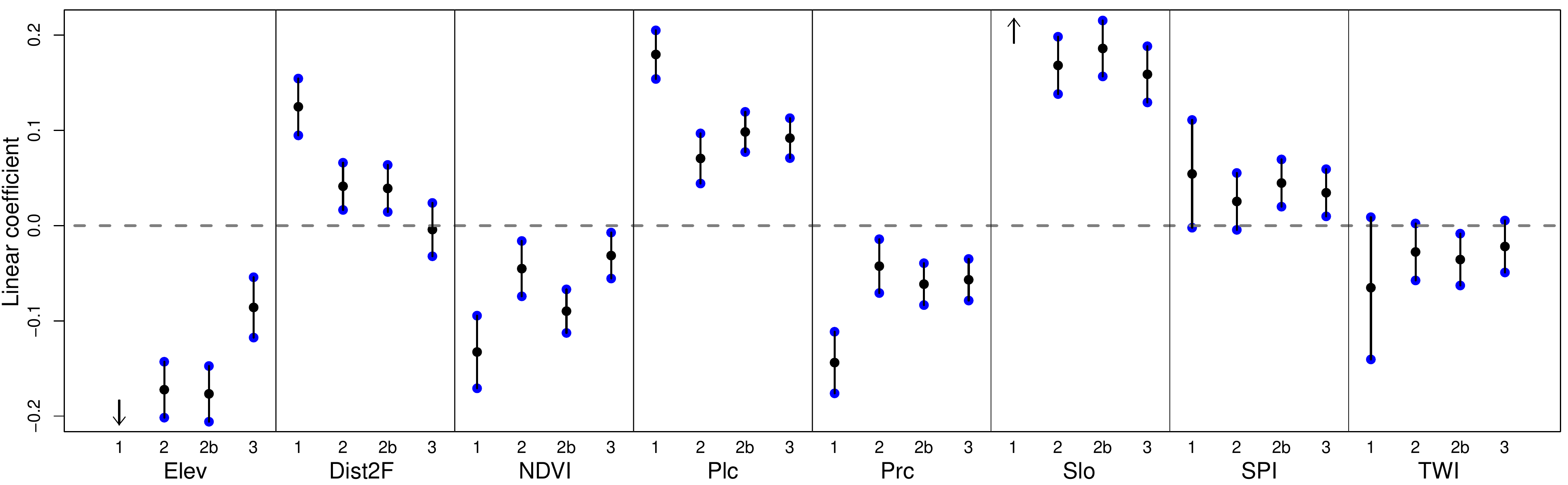}
	\caption{Posterior means (black dots) of fixed linear effects (except the Intercept) with $95\%$ credible intervals (vertical segments) for the Models 1, 2, 2b, 3. For Mod1, the coefficients of Elevation and Slope do not appear on the graph and are represented by arrows. Their posterior means ($95\%$ credible intervals) are equal to $-0.85$ $(-0.90,-0.80)$ and $0.75$ $(0.72,0.81)$, respectively. Acronyms of covariates can be found in Table~\ref{table1}.}
	\label{fig:fixedeffects}
\end{figure}

Figure \ref{fig:fixedeffects} shows the estimated fixed effect coefficients (except for the intercept whose interpretation is of minor interest here) for our four models. Recall that covariates have been rescaled to have mean $0$ and variance $1$. For better visualization, two estimates do not appear on this graph for Model Mod1: the coefficient for the Elevation has posterior mean $-0.85$ and $95\%$ credible interval $(-0.90,-0.80)$, while the Slope coefficient has posterior mean $0.75$ and $95\%$ credible interval $(0.72,0.81)$. Overall, a number of strongly positive and negative influences on landslide activation with relatively narrow credible intervals can be detected thanks to the large number of observed events and our fine spatial grid resolution. Estimated values are comparable across the most complex models, while the linear model Mod1 has several significantly higher coefficients in absolute value. An explanation is that part of these linear effects become part of the nonlinear and spatial effects in the other models; in particular, there may be some correlation between observed covariates on the one hand and unobserved covariates on the other hand, such as the intensity of the precipitation trigger. From a modeling point of view, the intensity variation may be better captured by the unobserved covariate effect rather than by the observed covariates. Moreover, the prior distribution of fixed and random effects may have a stronger influence in more complex models, yielding  slightly smoother posterior estimates. 

\begin{figure}[t!]
\centering
\includegraphics[width=.32\linewidth]{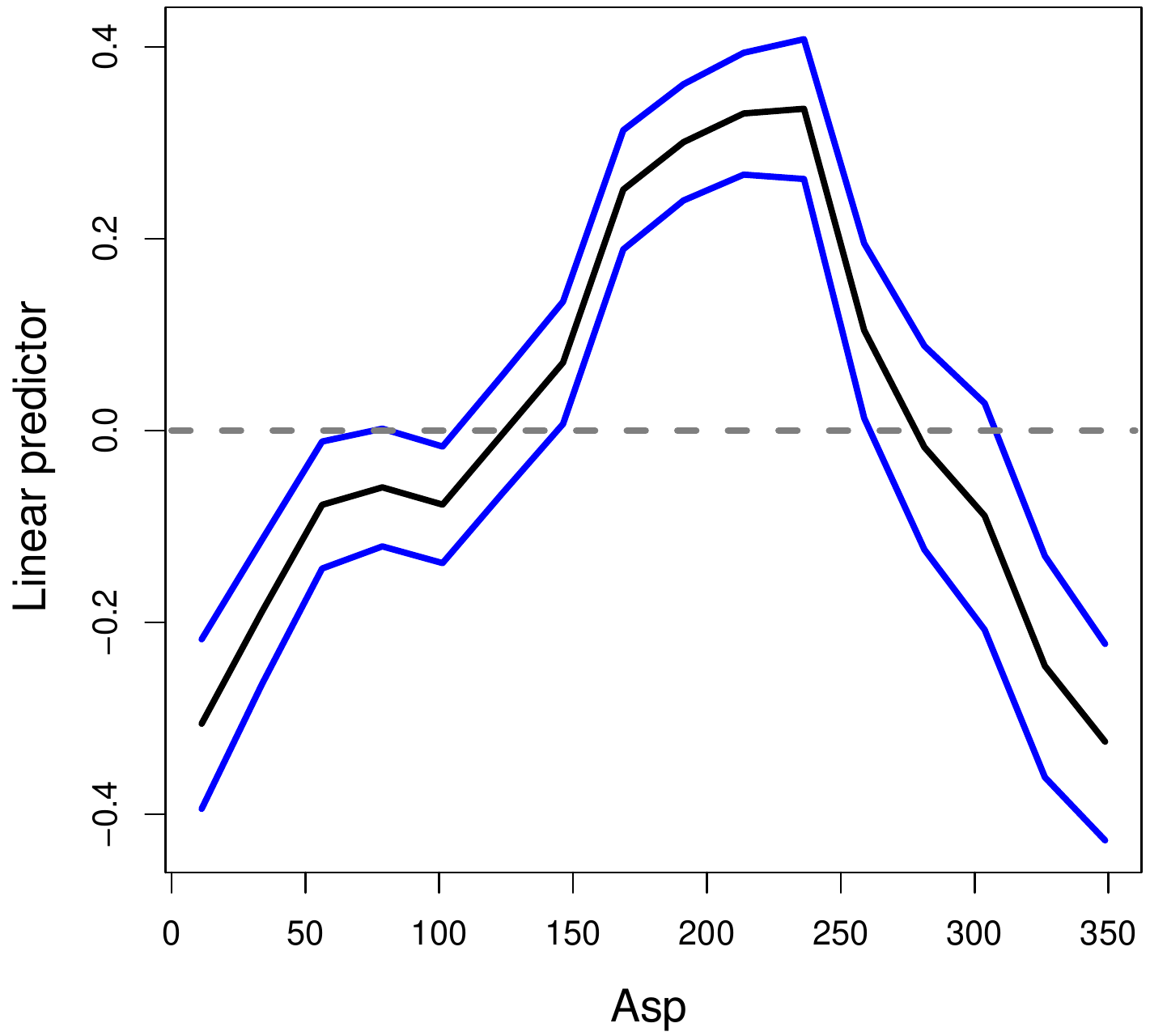}
\includegraphics[width=.32\linewidth]{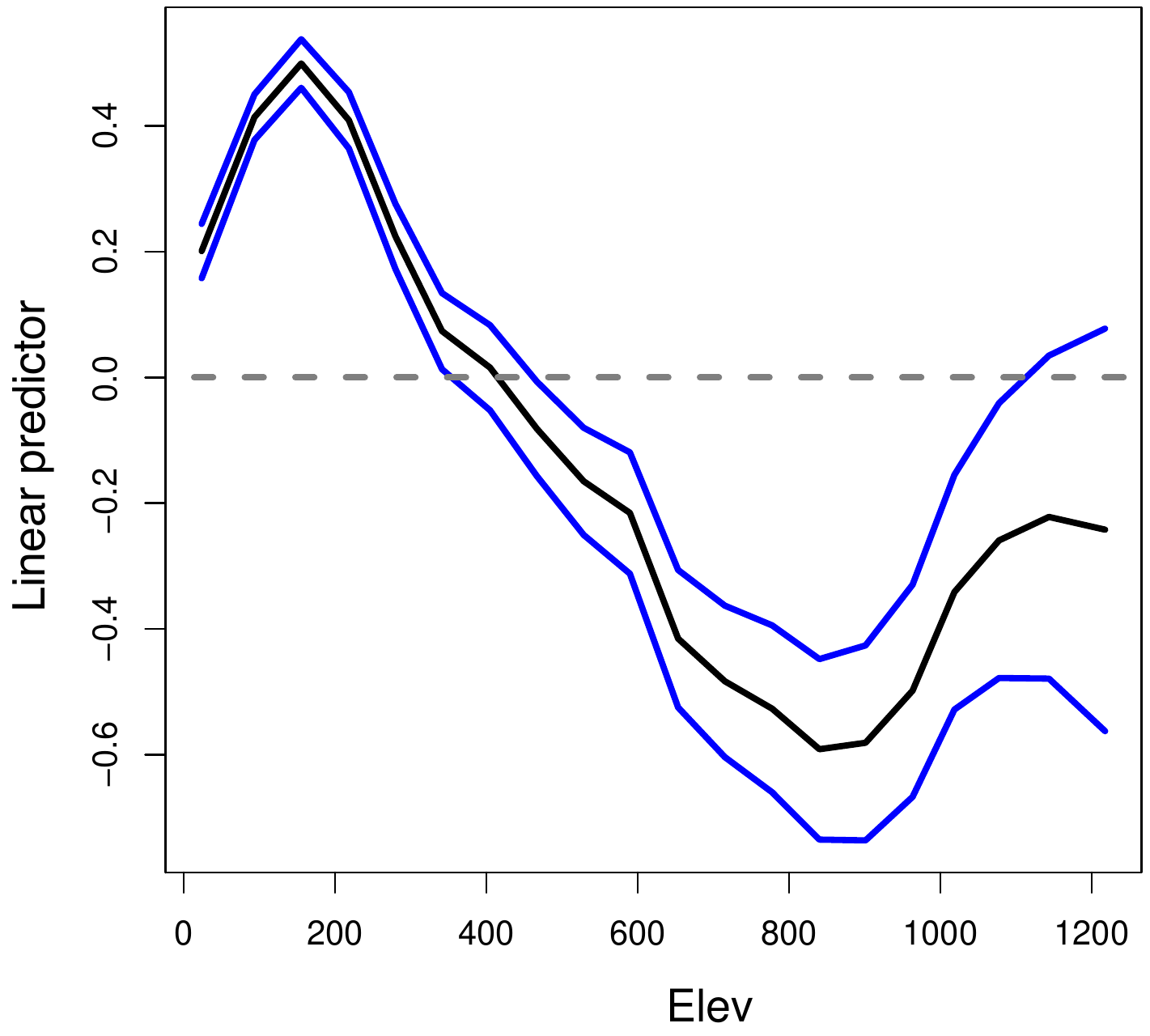}
\includegraphics[width=.32\linewidth]{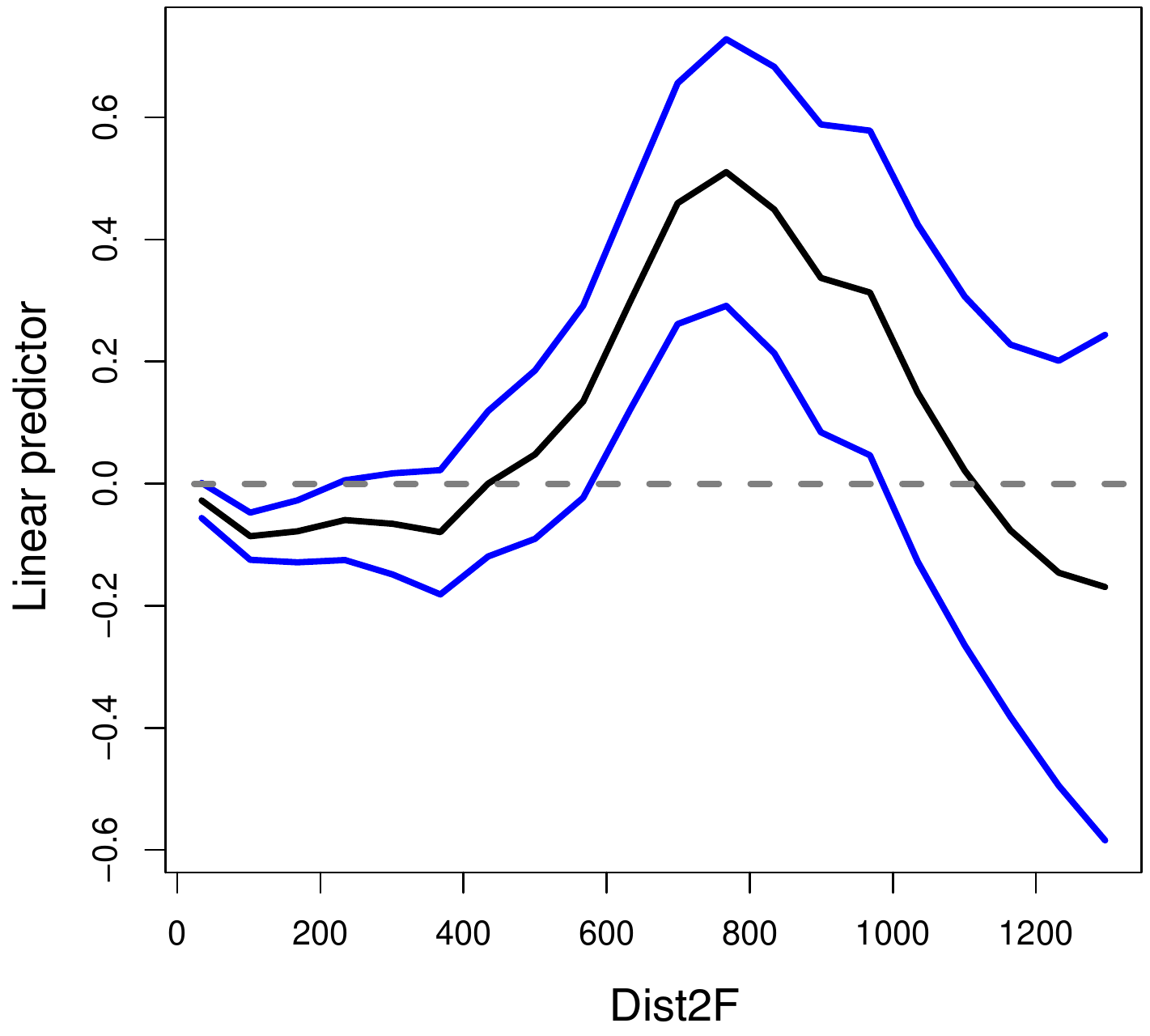} \\
\includegraphics[width=.32\linewidth]{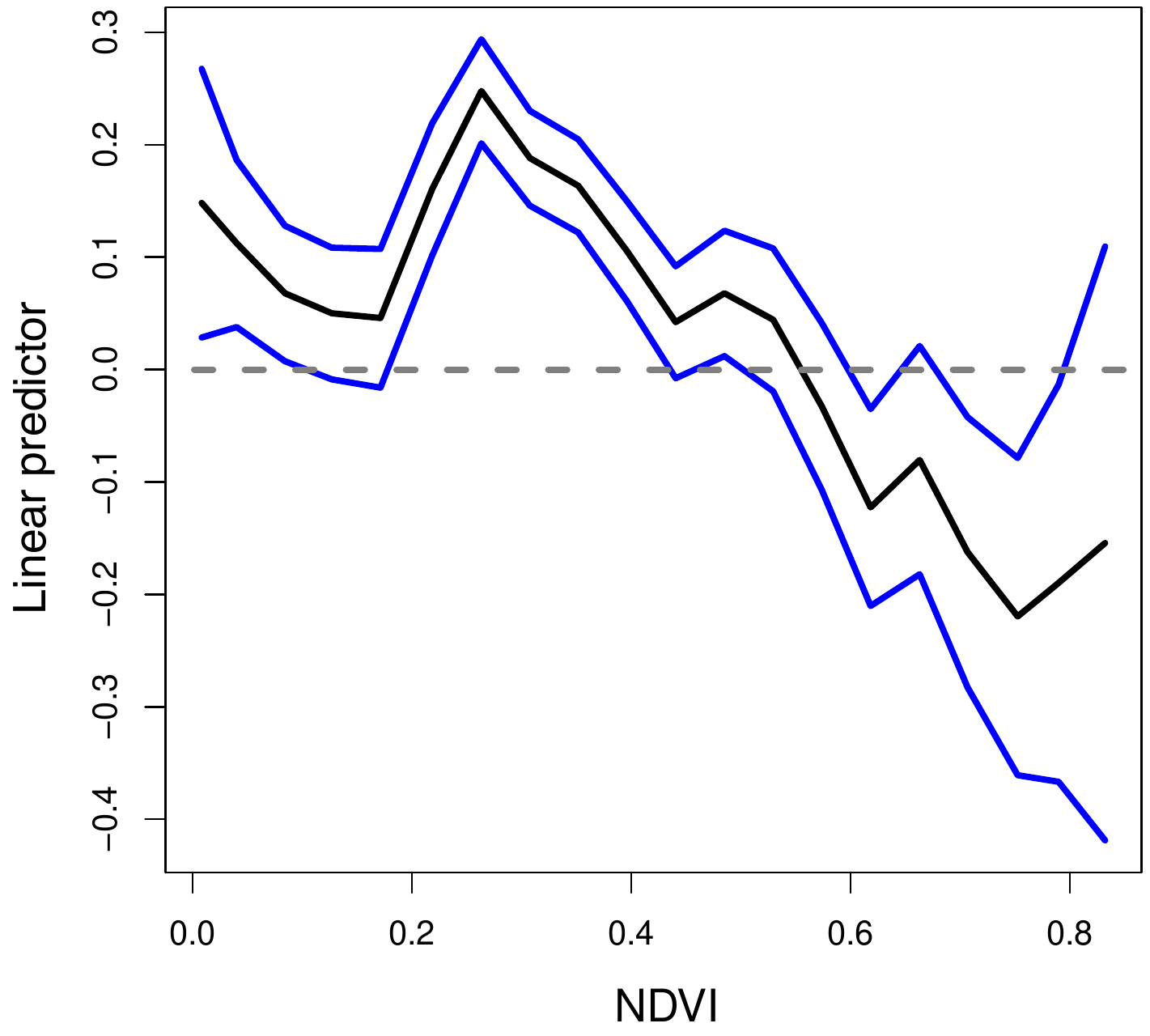}
\includegraphics[width=.32\linewidth]{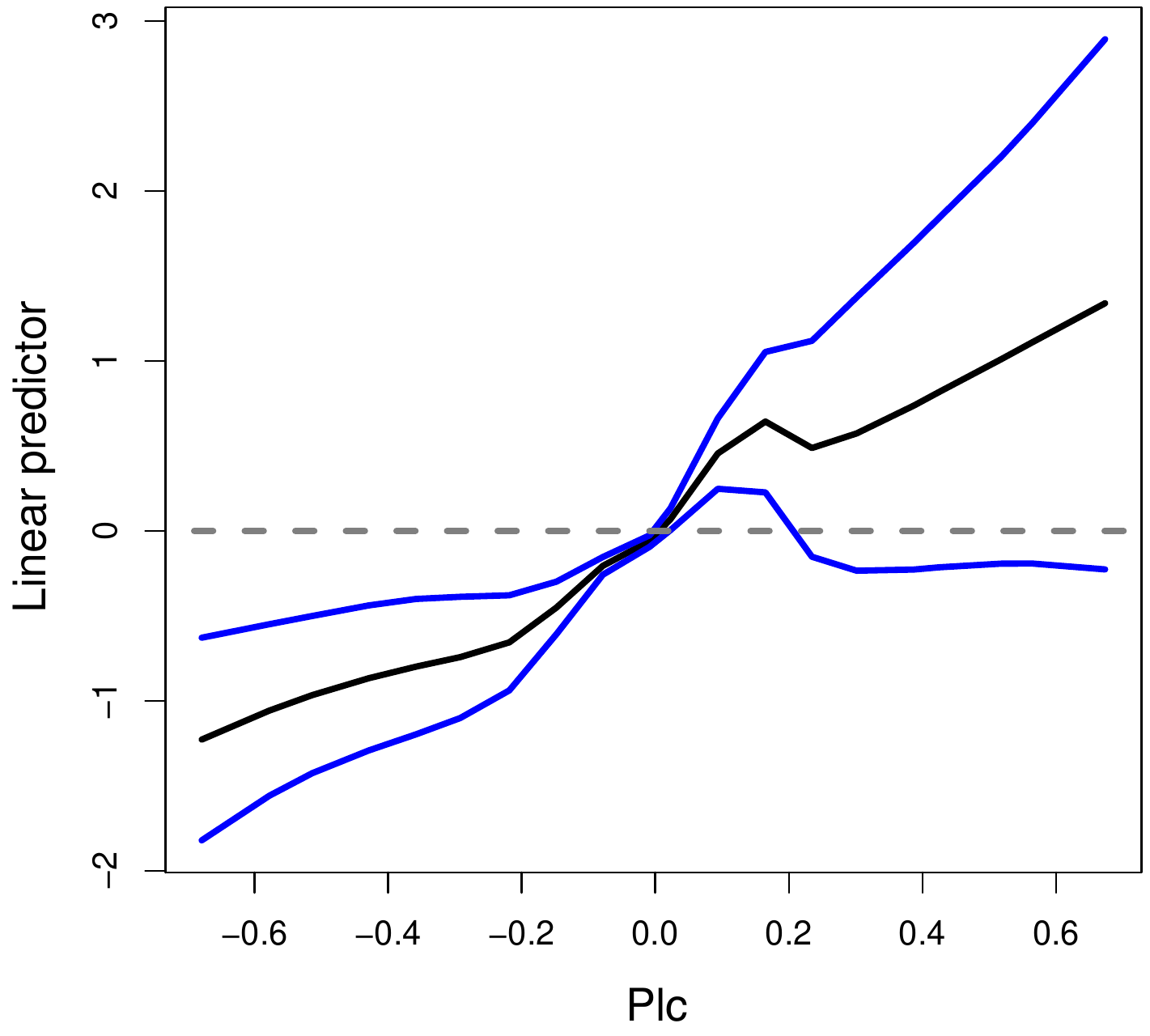}
\includegraphics[width=.32\linewidth]{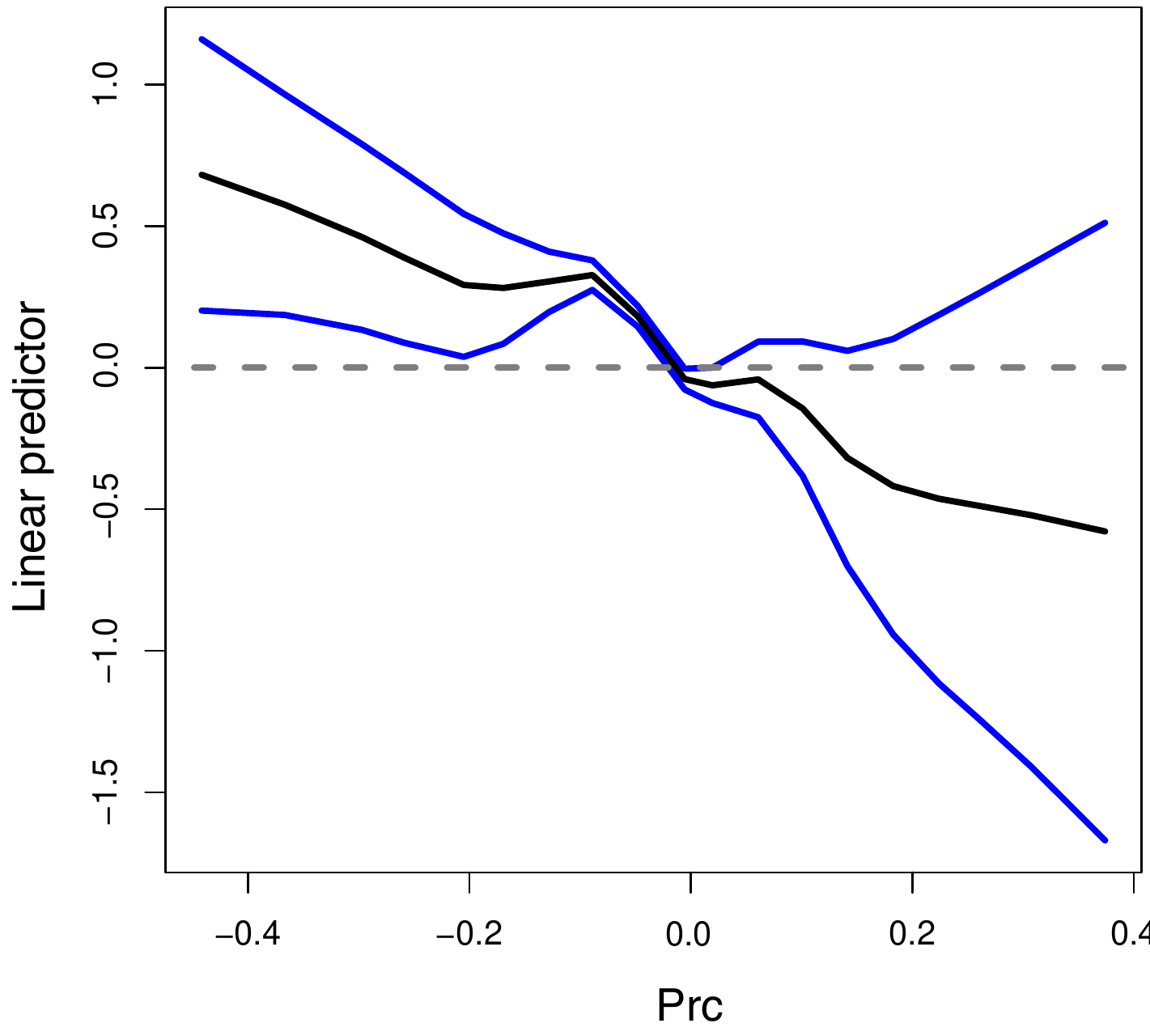} \\
\includegraphics[width=.32\linewidth]{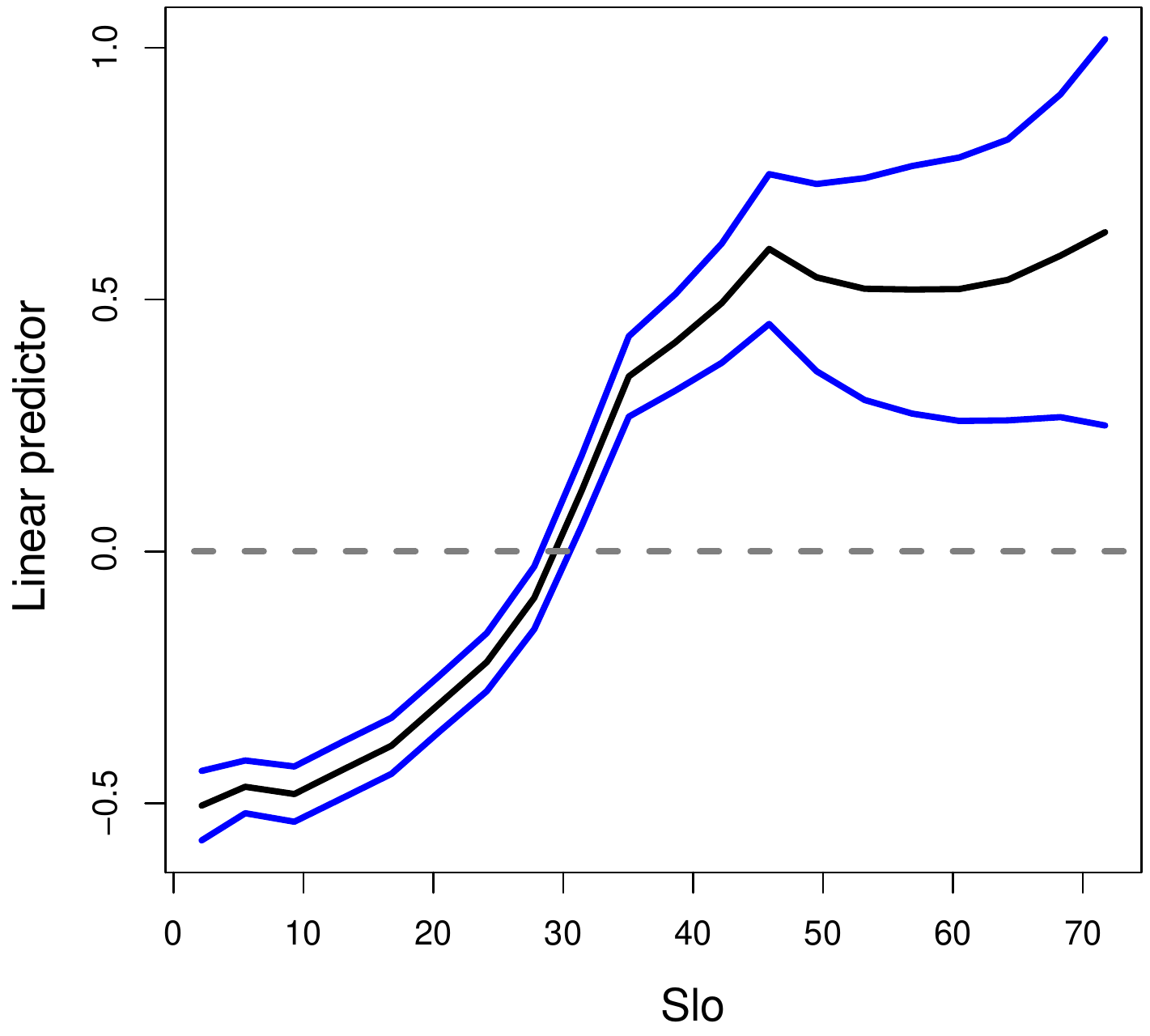}
\includegraphics[width=.32\linewidth]{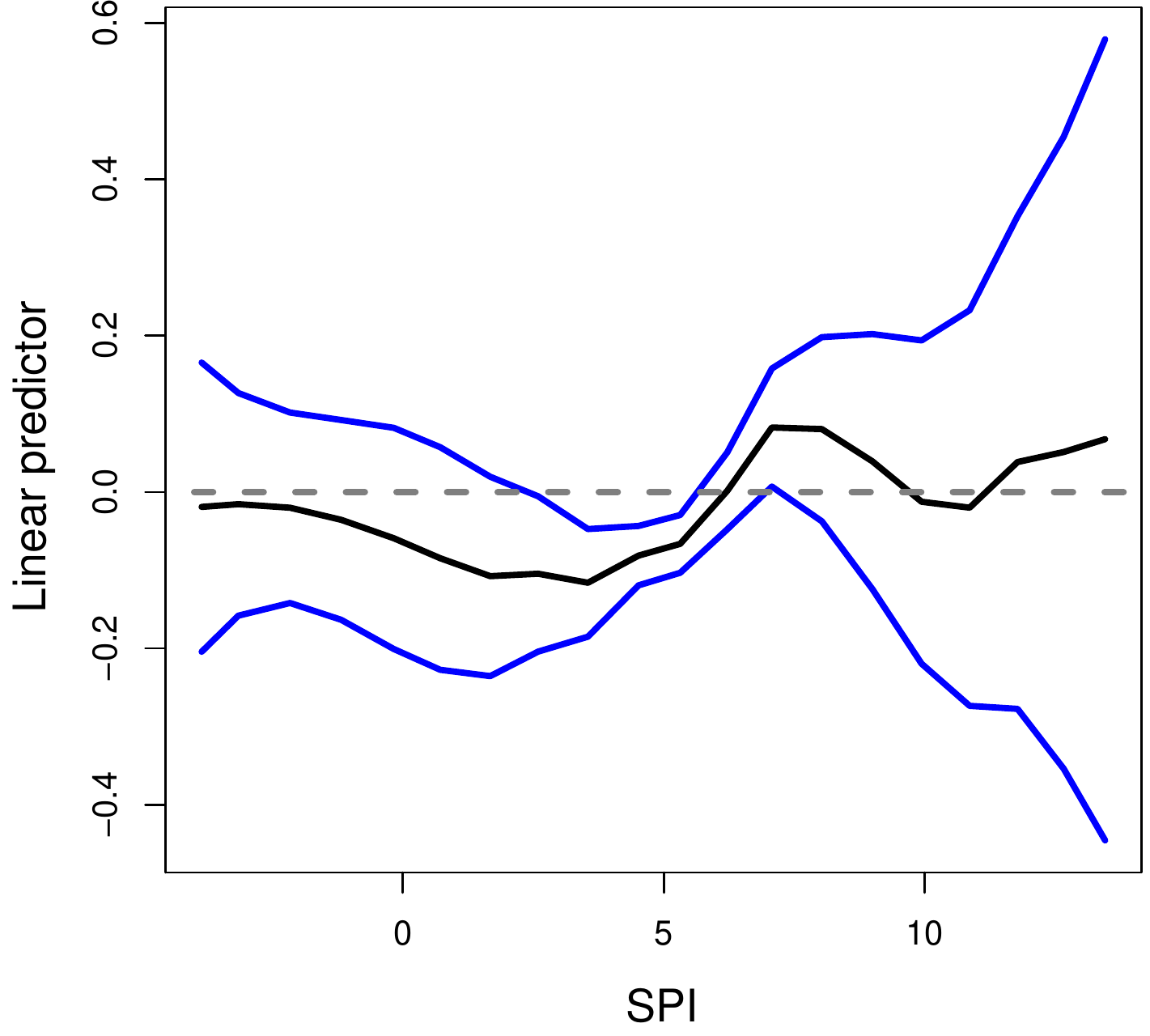}
\includegraphics[width=.32\linewidth]{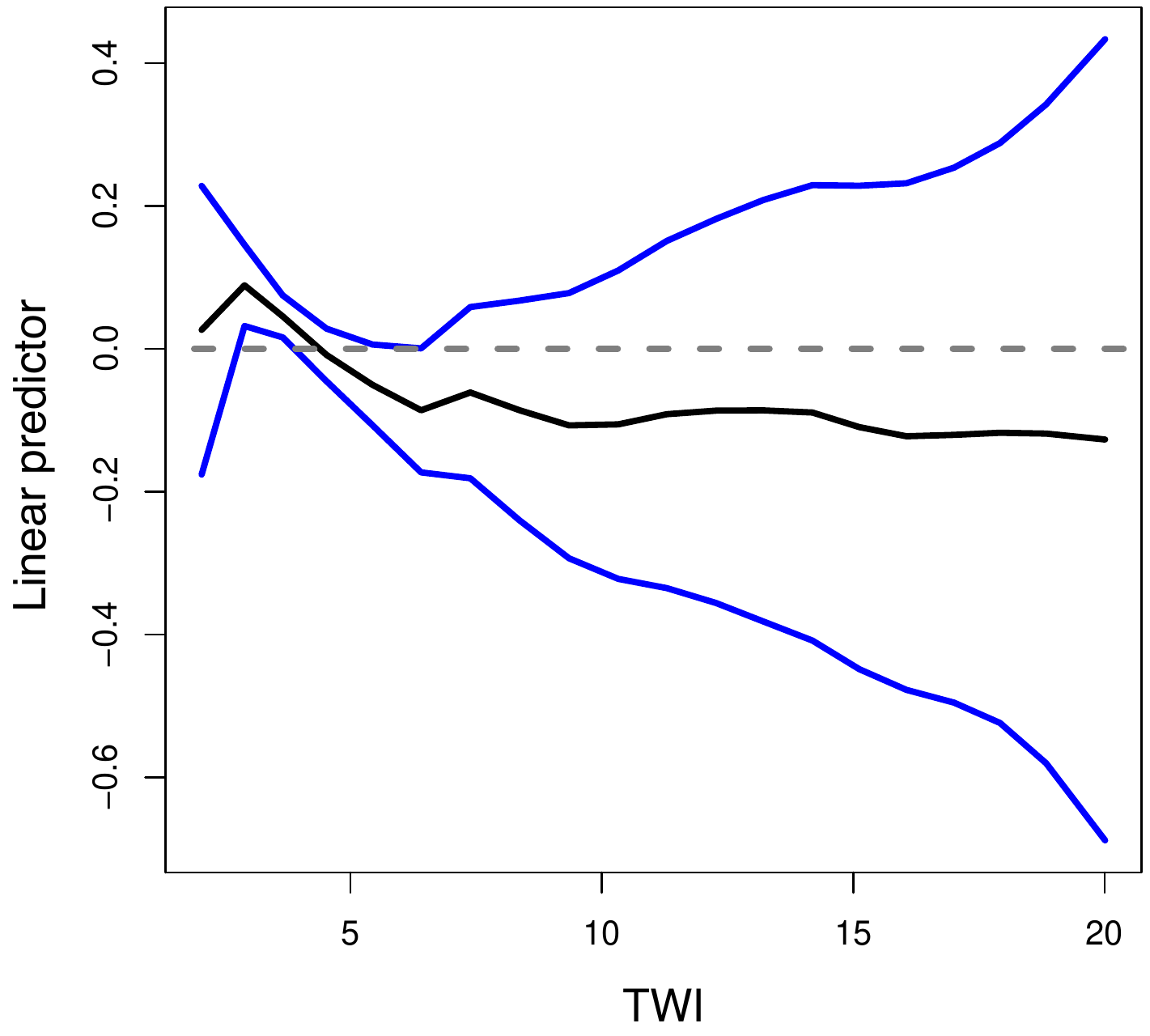}
\caption{Pointwise posterior means (black curves) and $95\%$ credible envelopes (blue curves) for the overall (linear and nonlinear) continuous covariate effects in Model Mod2. Acronyms of covariates can be found in Table~\ref{table1}.}
\label{fig:mod2}
\end{figure}

We now consider Model Mod2, in which all continuous covariates are treated nonlinearly. Figure~\ref{fig:mod2} displays the overall covariate effect (linear plus nonlinear) with estimated $95\%$ credible intervals; a fully linear model would yield straight lines, while this more flexible nonlinear model allows for departures from this idealistic situation. Covariate values where all three lines (posterior estimate and credible envelopes) lie above $0$ indicate a significant positive contribution to  landslide intensity, and by symmetry values below $0$ of all three lines correspond to significant negative effects. 
The Aspect has a strongly nonlinear effect; broadly speaking, SSW-facing slopes are the most prone to landsliding and N-facing ones are the least prone, with a smooth transition in between. Elevation has an overall negative correlation with landsliding with the exception of the range from $0$ to $200$m.a.s.l. (positive effect of increasing strength)  and above $900$m.a.s.l. (negative effect of decreasing strength). 
The significant positive effect of the Distance to Fault lines peaks at approximately $700$m, whilst it does not seem to be important up to $500$m and beyond $1000$m. NDVI is quite noisy, but it still appears to be negatively correlated with landsliding overall, with lower NDVI values having slightly stronger effects. 
Planar and profile curvatures are clearly positively and negatively correlated with landslides, respectively, and they do no show any significant departures from linearity. In particular, sidewardly (respectively upwardly) convex morphologies are more prone to landsliding than concave ones. Slope steepness is among the most significant effects with very narrow credible intervals up to roughly $50\degree$. As expected, steeper slopes are more at risk. The Slope effect appears to be highly nonlinear with increasingly unstable conditions up to $50\degree$, beyond which the Slope effect tends to stabilize although with greater uncertainty due to the lack of observations in this range of slope values. The Stream Power Index (SPI) effect is quite irregular and does not seem to be significant overall, although it is slightly stronger for $\mbox{SPI}>6$. Finally, the Topographic Wetness Index (TWI) effect is not significant overall, although smaller TWI values appear more strongly correlated with landsliding. 

\begin{figure}[t!]
	\centering
	\hspace{-5pt}\includegraphics[width=\linewidth]{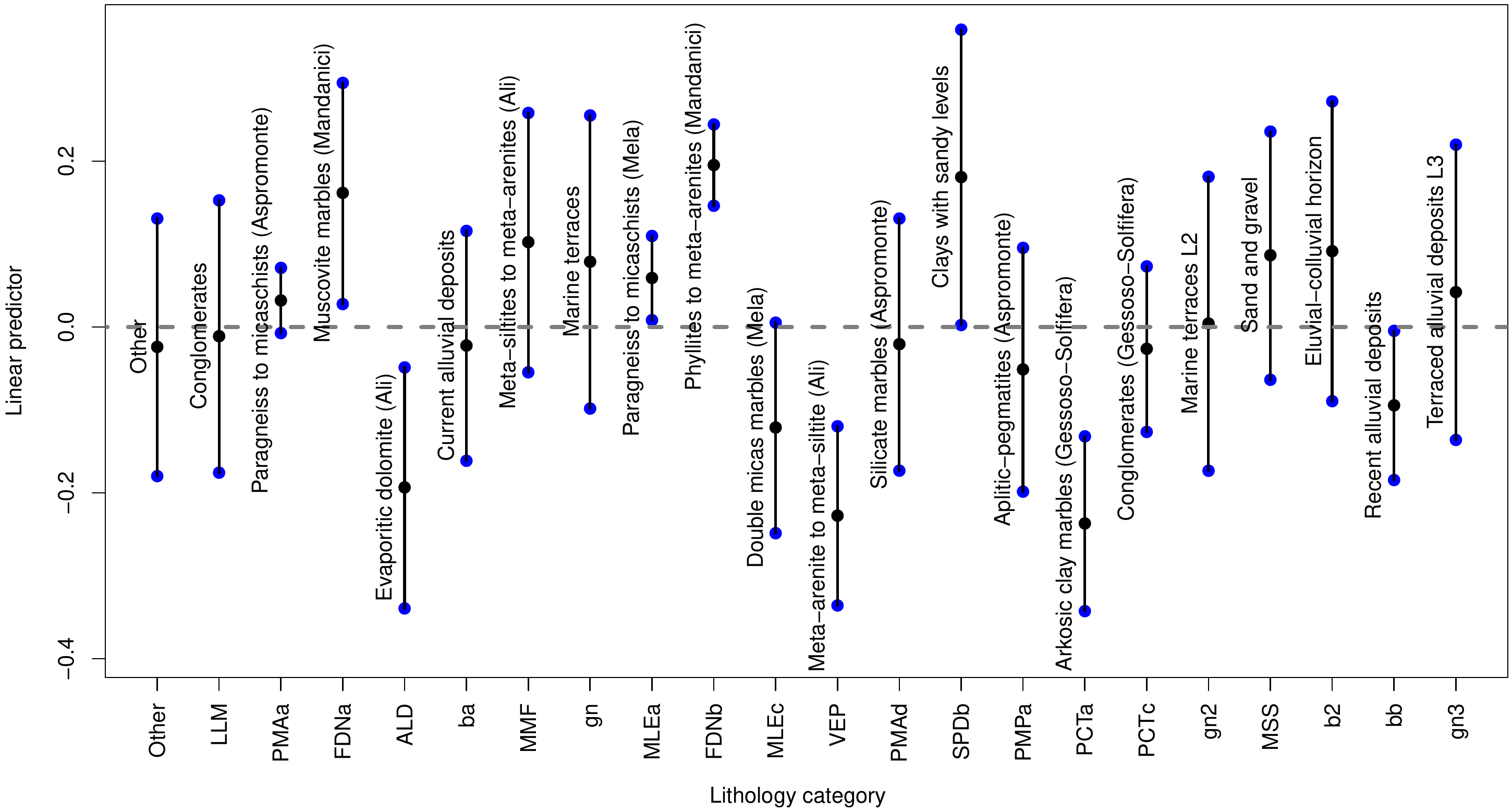}\\
	\vspace{5pt}
	\includegraphics[width=0.49\linewidth]{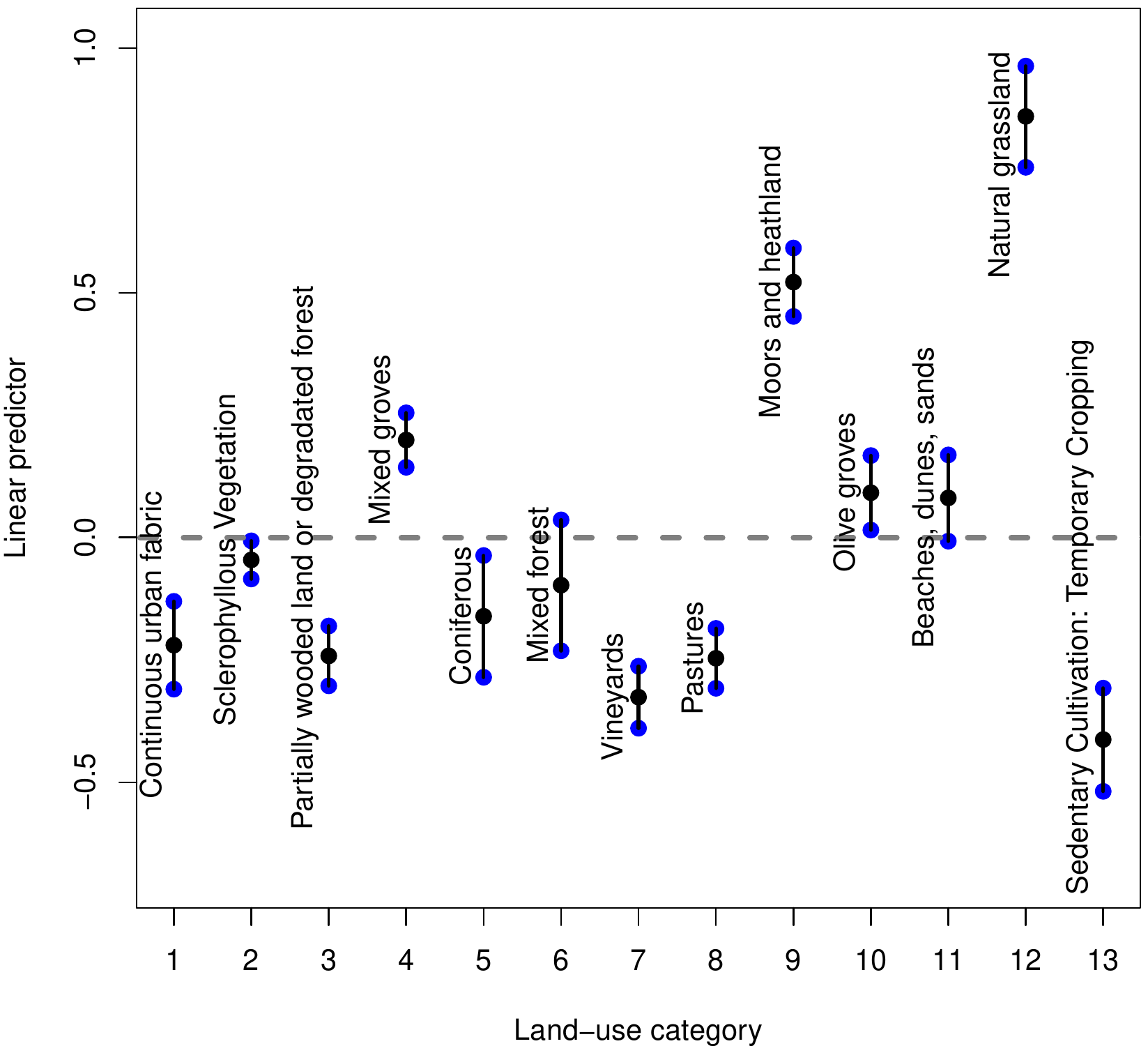}
	\includegraphics[width=0.49\linewidth]{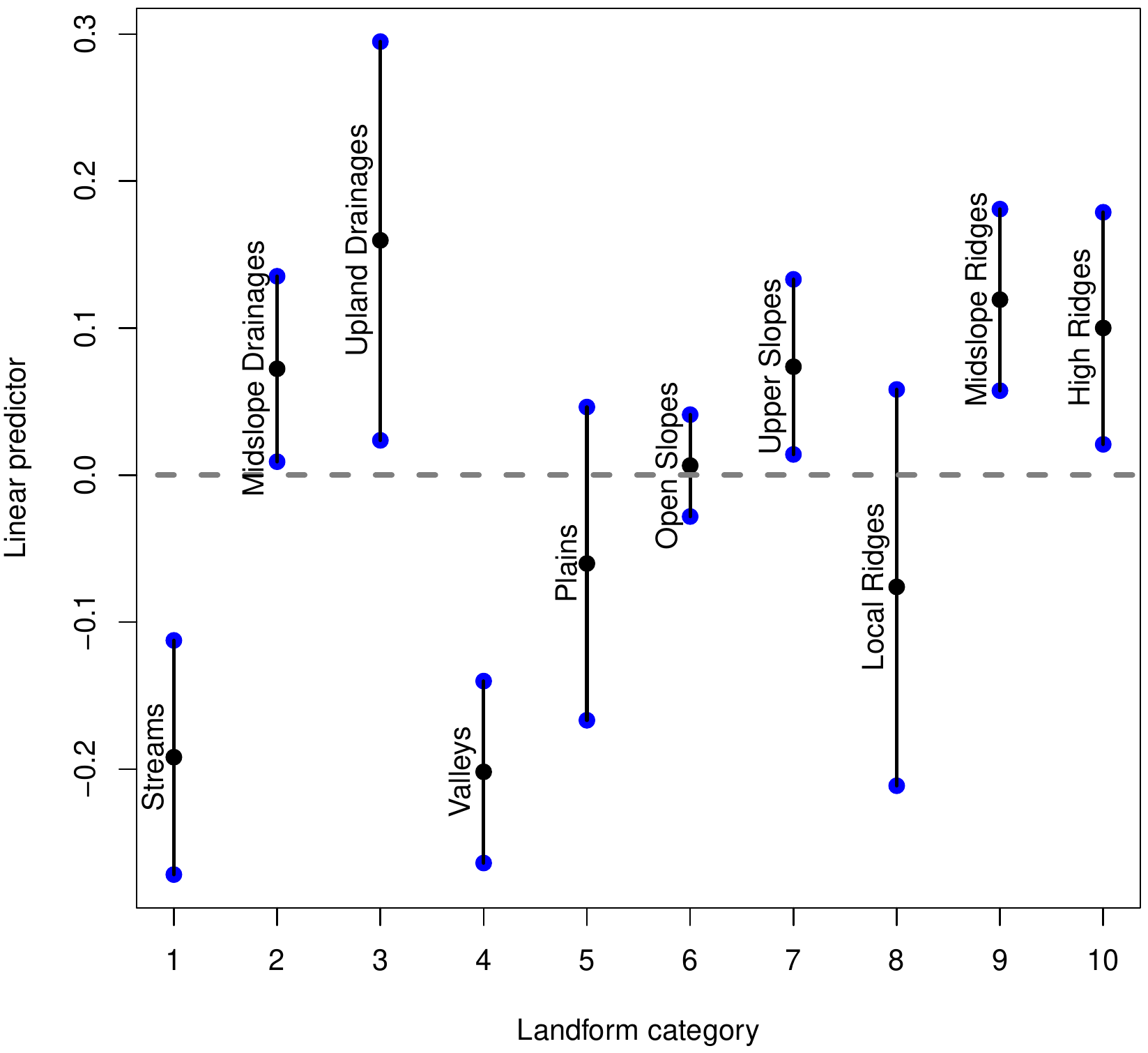}
	\caption{Posterior means (black dots) of estimated effects, with $95\%$ credible intervals (vertical segments) based on Model Mod2b, for Lithology (top), Land Use (bottom left) and Land Form (bottom right). Lithology categories with very few occurrences in the study region are summarized in ``Other".}
	\label{fig:litho.land}
\end{figure}

To simplify the model and achieve more efficient inference, we now examine the significance of nonlinear effects in Mod2, illustrated in Figure~\ref{fig:mod2}. Based on statistical and geomorphological arguments, we finally retain Aspect, Elevation, Distance to Faults and Slope as the only nonlinear covariates in Model Mod2b. The remaining continuous covariates are treated linearly. As the results were quantitatively the same for Mod2 and Mod2b, we do not report all results for the simplified model. However, as suggested by Figure~\ref{fig:fixedeffects}, estimated fixed effects in Mod2b are comparable to Mod2, but they tend to be slightly stronger with shorter credible intervals. Figure \ref{fig:litho.land} shows estimates and $95\%$ credible intervals for all non-ordinal categorical covariate effects in Mod2b. Amongst $22$ Lithology classes $8$ levels appear to be significant, together with $11$ levels  out of the $13$ Land Use classes and $7$ levels out of $10$ Land Form classes. A considerable number of other classes miss significance by little. 

In our last model Mod3, we keep the same selection of nonlinear covariates as in Mod2b, but we now also include a latent spatial effect to account for unexplained variations in the landslide intensity function. The results for linear and nonlinear effects are generally quite similar to those obtained for Mod2 and Mod2b (recall Figure~\ref{fig:fixedeffects} for fixed linear effects), although certain covariates (such as Distance to Faults for example) are ``absorbed'' into the latent spatial effect, and therefore lose significance in Mod3. Section~\ref{sec:SpatialEffect} further describes the estimated spatial effect.

\subsection{Characterization of the estimated latent spatial effect}\label{sec:SpatialEffect}
\begin{figure}[t!]
	\centering
	\includegraphics[width=\textwidth]{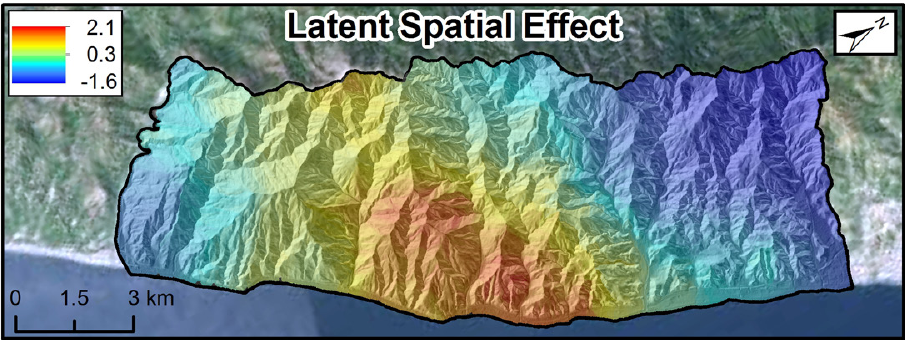}
	\caption{Posterior mean of the slope-unit based latent spatial effect in Model Mod3.}
	\label{fig:SpatialEffect}
\end{figure}
Figure \ref{fig:SpatialEffect} shows the posterior mean surface of the latent spatial effect $W_{\bm z_1}(s)$ defined through \eqref{eq:spateff} and used in our model Mod3. The fitted spatial effect is piecewise constant over slope units. Spatial dependence is strong, as estimated values vary quite smoothly over space. The highest values are observed near the coast in the middle of the study region, where lots of landslides occurred, while lower values are found in the Northern subregion. An obvious negative gradient is visible when moving away from the epicentral area. Unreported results show that the estimated spatial effect is clearly significant in terms of slope unit-wise credible intervals in the high and low value regions, implying that the spatial effect as a whole is extremely significant and therefore plays a crucial role in explaining the spatial variability of the landslide intensity function. The spatial precision parameter $\tau_1$ in the CAR model, recall \eqref{eq:spateff}, has posterior mean $2.7$ with a relatively narrow $95\%$ credible interval given by $[2.3,3.3]$, which confirms that neighboring slope units are strongly correlated. Compared to the simpler models (Mod1, Mod2 and Mod2b) in which only the observed covariate effects are present, this spatial effect modifies the intensity to have values that are up to $5$ times lower (minimum spatial effect value $-1.6$) or up to $8$ times higher (maximum spatial effect value $2.1$). It therefore corrects for the strong over-dispersion of counts in the Poisson process models without spatial effect, and it can capture complex clustering patterns.

\subsection{Predictive performance and comparison of models}

\begin{figure}[t!]
	\centering
	\includegraphics[width=\linewidth]{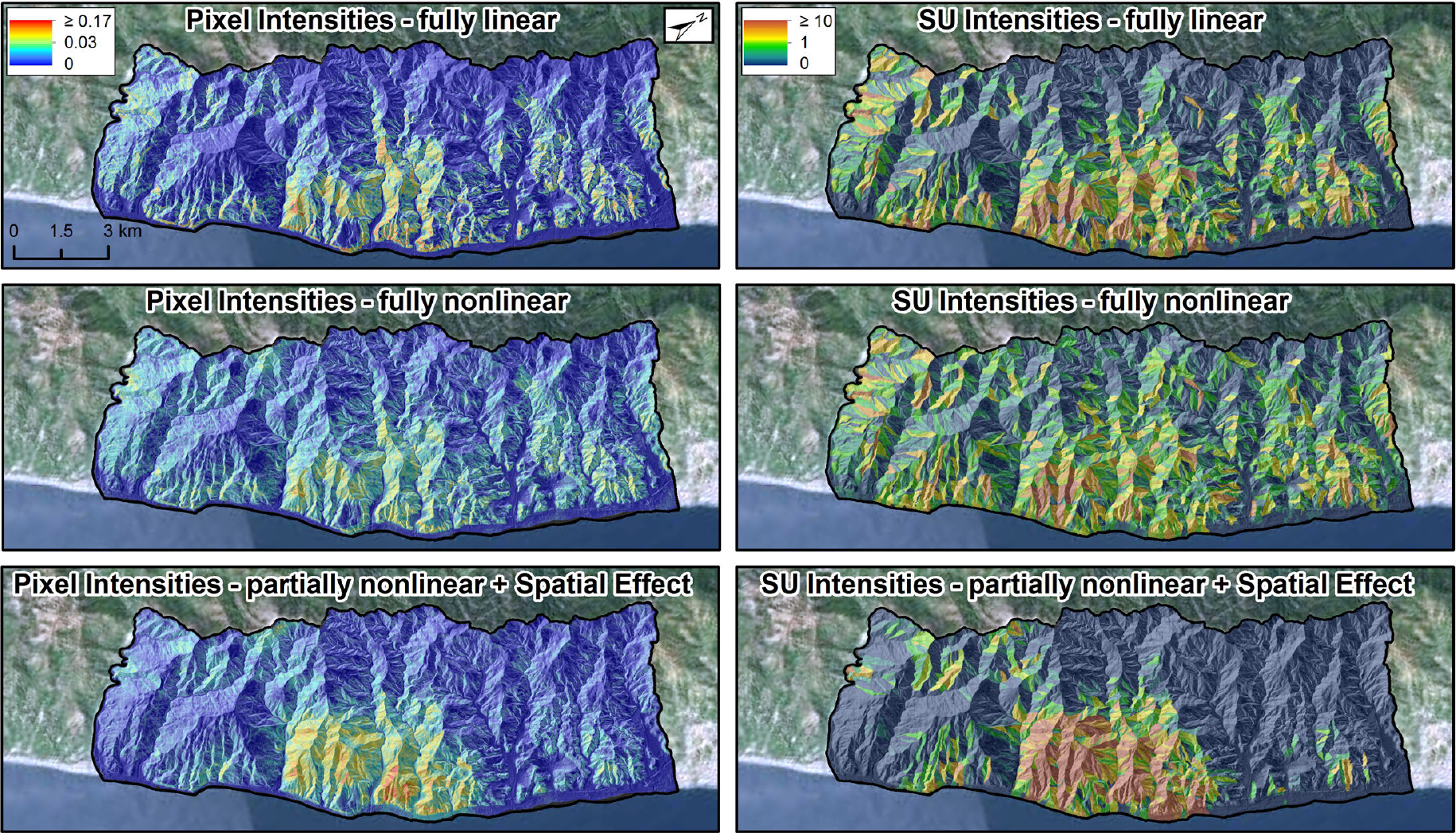}
	\caption{Predicted intensity maps for Models Mod1 (top), Mod2 (middle) and Mod3 (bottom), using a spatial resolution based on pixels (left) or an aggregation to slope units (right).}
	\label{fig:intensity}
\end{figure}

Figure~\ref{fig:intensity} displays the predicted intensities for Models Mod1, Mod2 and Mod3 on the pixel level, and aggregated to the slope unit scale. Recall that the intensity may be interpreted, up to a constant, as the expected number of landslides likely to occur at a specific location. While the estimated intensity function appears to be fairly similar for Mod1 and Mod2, there are significant differences with Mod3. The presence of the spatial effect in Mod3 allows to capture abrupt variations in the intensity function and is able to simultaneously estimate very high and very low intensities in different regions of space. Without the spatial effect, Mod1 and Mod2 are more rigid and do not adequately represent the complex landslide point pattern characterized by a high density of events located in the epicentral area and sparse events in the marginal sectors (recall Figure~\ref{fig:map}(d)).

\begin{figure}[t!]
	\centering
	\includegraphics[width=0.6\textwidth]{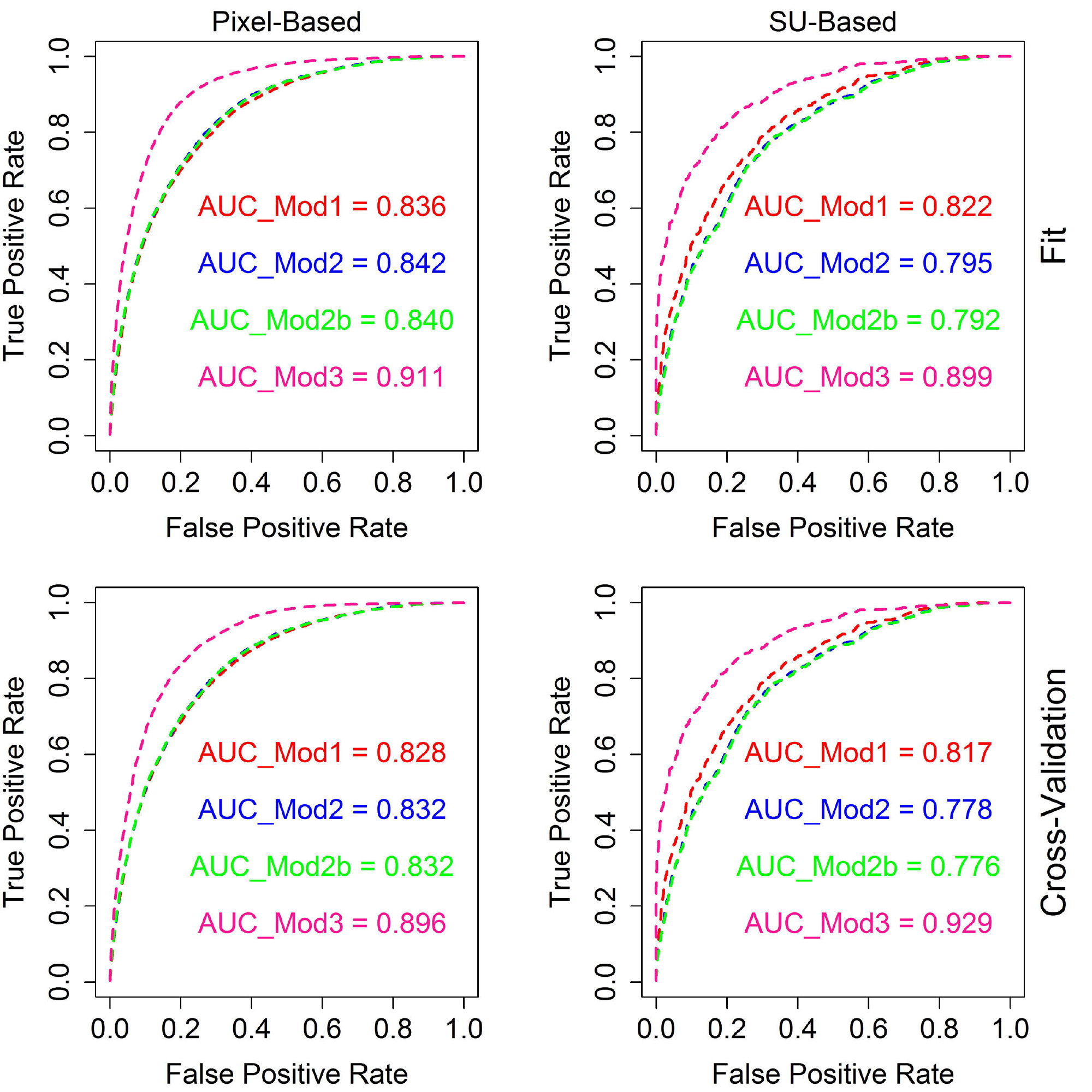}
	\caption{ROC curves and corresponding AUC values for Models Mod1 (red), Mod2 (blue), Mod2b (green) and Mod3 (pink). These performances are displayed for the actual fit (top row) and the cross-validation (bottom row), on pixel (left) and slope unit (right) levels.}
	\label{fig:performances}
\end{figure}

To assess the \emph{within-sample} performance of each model, we then computed the fitted probabilities of observing at least one landslide in each pixel and slope unit in the study region, using the formulae \eqref{eq:pi} and \eqref{eq:aggr}. 
We computed the Receiver Operating Characteristic (ROC) curve, as well as the Area Under the Curve (AUC), by comparing fitted values with observed data. The latter are the most common performance metrics used in geomorphological studies. The larger the AUC the better the fit: for $\mbox{AUC}=0.5$, the fitted model arbitrarily discriminates between stable and unstable conditions, while for $\mbox{AUC}=1$, the model fits the data perfectly. We did similar calculations to check the \emph{predictive} (i.e., \emph{out-of-sample}) performance of each model. More precisely, we conducted a $4$-fold cross-validation experiment by removing approximately $1/4$ of the slope units (randomly selected, but each slope unit removed only once) from the data. 
Figure~\ref{fig:performances} summarizes the results.

The models without spatial effect have excellent performances according to \citet{hosmer2000} with AUC values slightly higher for pixel-based predictions ($0.836 < \mbox{AUC} < 0.842$) than for slope units ($0.792 < \mbox{AUC} < 0.822$). The same pattern arises in the cross-validation results for the models without spatial effect, with pixel-based AUC values ranging between $0.828$ and $0.832$, whereas the slope unit partitioning yields a stronger drop in prediction skill ($0.776 < \mbox{AUC} < 0.817$). Prediction results for the spatial effect model Mod3 are outstanding, reaching an AUC of $0.929$ for the cross-validation based on slope units, and clearly outperform the simpler models. This level of accuracy is unprecedented in the geomorphology literature. Moreover, the gap between pixel- and slope unit-based performances is extenuated thanks to the inclusion of the spatial latent effect. 
Our good results for both within-sample and out-of-sample prediction experiments confirm that our models fit the data well, while avoiding overfitting, which is crucial for reliable risk assessment.

\section{Geomorphological interpretation of statistical results}
\label{sec:interpretation}
\subsection{Summary of modeling novelties}
Our new technique assesses the landslide susceptibility to activation for multiple debris flow. Beyond estimating the spatial probability of landslide occurrence, we also predict the potential number of landslides per (arbitrary) unit area resulting from an extreme storm. We also integrated a linear versus nonlinear evaluation of causative predictors and the extraction of a latent spatial  effect from the data. We now provide a more detailed geomorphological interpretation of the statistical results described in Section~\ref{sec:estimation}.

\subsection{Linear versus nonlinear effects in continuous covariates}
Linear and nonlinear covariate effects have a tendency to show a general agreement concerning the overall trend, but the nonlinear implementation allows for a more detailed interpretation and reveals some important nonlinear effects. In particular, the Aspect variable is strongly correlated to landslides in SE, S, SW and W directions, in agreement with other studies of the area. \cite{cama2017improving} obtained negative regression coefficients for Eastness and Northness indicating West and South as preferential instability directions. \cite{lombardo2016a} show an increase in landslide susceptibility in SE, S, SW directions; however, slight dissimilarities may be due to the study area they investigated, limited to the catchment of Mili located in the Northern part of our study area.

Many papers covering the studied event \citep[e.g.,][]{lombardo2014test,cama2015predicting}  have shown that Elevation is negatively correlated to landslides, although it is often included into models as a proxy for precipitation. The latent unobserved effect displayed in Figure \ref{fig:SpatialEffect} provides good insight into the rainfall pattern that might have occurred during the Messina disaster, while elevation is shown to have a positive and increasingly strong effect  up to $200$m.a.s.l. and then decreases to reach a relatively negative effect. This may be due to the storm dissipating its discharge at lower altitudes without reaching the catchment ridges. Similar hypotheses and related plots are given in \cite{lombardo2015binary} but limited to the epicentral catchments of Giampilieri and Briga.

Distance to fault lines contributes to slope instability at relatively large distances from around $500$m to $1000$m, with a peak at $700$m. We interpret this  as being due to the absence of the tectonic effect for near distances where tectonized materials are easily weathered and then already removed prior to the landslide occurrences through common erosional processes. Conversely, very far from the fault lines the mantle draping over the bedrock is less susceptible to average climatic stresses and requires a greater rainfall discharge to mobilize and evolve into debris flows. A similar conclusion was observed by \cite{lombardo2016a}. However, the latter estimated a slightly shorter distance to fault lines up to which the probability of landslide is positive (the maximum effect was roughly estimated to be $600$m).

The NDVI, which captures the density of the vegetation, negatively contributes to the intensity of landslides. This result agrees with the general assumption  \citep[e.g.,][]{elkadiri2014remote} that the bare soil is more directly exposed to the trigger when the vegetation is sparser. 

Slope steepness positively affects landslide intensities, especially from $28\degree$ to $70\degree$. This result is in good agreement with the general assumption that debris flows initiate in channels steeper than $20\degree$ \citep{imaizumi2006hydrogeomorphic}.

The SPI effect is not significant overall, but appears slightly positive from values greater then $6$. As SPI is a proxy for runoff erosive power, the increasing trend from low to high SPI values appears to be reasonable and in agreement with other contributions correlating high SPI values with greater landslide occurrences \citep{devkota2013landslide}.

The TWI effect is not significant but is markedly negative beyond the value of $4$. The effect appears to be slightly positively correlated at very low TWI values, followed by a smooth attenuation towards topographic conditions more prone to retain water (i.e., plains). This trend agrees well with other papers where the same pattern is shown in landslide frequencies \citep{pourghasemi2012landslide}.   

\subsection{Categorical covariates}
The categorical covariates in the nonlinear model have slightly different coefficient estimates compared to the linear one, see Figure \ref{fig:litho.land}, which we attribute to the higher model complexity in the nonlinear model leading to some loss in statistical power for detecting significance of effects. \textit{Recent Alluvial Deposits} are not included among the significant classes for the nonlinear model, despite the  significant negative contribution in the linear model. Similarly, significant linear Landforms include \textit{Midslope drainage}, \textit{Plains} and \textit{High ridges}, which are not significant in the nonlinear counterpart. However, their effects appear to be consistent across models, namely positive, negative and positive, respectively, both in the linear and nonlinear cases. Consistency can also be found between estimated significant Land Uses in terms of class signs. However, there are still a few differences between linear and nonlinear models. For example, the linear model finds \textit{Olive groves}, \textit{Beach, dunes and sand}, and \textit{Shrubland} to be significant, whereas the nonlinear one finds \textit{Coniferous} and \textit{Natural Grassland}. 

From an interpretative perspective, \textit{Muscovite Marbles}, \textit{Paragneiss to Micaschists} and \textit{Phyllites to meta-arenites} correspond to metamorphic lithotypes known to be strongly weathered in the area \citep{lombardo2014test} making them unstable under strong meteorological stresses. As for \textit{Mixed groves}, \textit{Moors and heatland} and \textit{Natural grassland}, these land uses correspond to areas with low vegetation density. This condition offers a greater soil exposure to the impact of rainfall and hence to erosion. Similar results are shown for Southern Italy \citep[e.g.,][]{Pisano2017} and are generally emphasized by abandonment of agricultural practices. Conversely, \textit{Vineyards}, \textit{Pastures} and \textit{Temporary cropping} imply active care and constant control by local farmers, which plays a counter-effect to landslide activations. \textit{Degradated Forest}  is shown to behave similarly, which is in line with the considerations of \cite{reichenbach2014influence} in the catchment of Briga, where forested areas have been shown to favor terrain stability. 
Finally, \textit{Upland Drainages}, \textit{Upper Slopes} and \textit{Midslope Ridges} play a positive role to landsliding. These landforms correspond to steep or very steep portions of the landscape where instabilities take place due to gravitational processes.    

\subsection{Spatial intensity and precipitation trigger}
The predicted landslide intensity is shown in Figure~\ref{fig:intensity}. Spatial patterns  coincide with those already stated in the literature. Out of the twelve major catchments investigated in the present contribution, only a subset has been taken into consideration in other studies. In particular, Giampilieri has been assessed several times in terms of landslide susceptibility. \cite{deguidi2013} used the index method proposed by \cite{van1997statistical} producing analogous predictive patterns in the left bank. Similarly, the Briga catchment is depicted as highly susceptible both in the left and right flanks in \cite{rossi2016land} based on logistic regression techniques, which is also the case for predicted intensities in the present work.
Further analogies can be drawn between the current results for the catchments of Itala and Mili with \cite{cama2017improving} and \cite{zini2015rusle}, respectively. 
However, the introduction of the latent spatial effect changes the spatial predictive patterns, differentiating them from those compared above. The intensities are shown to focus in the epicentral sector where the climatic stress has reached its maximum discharge; moreover, this increases the predictive performance, recall Figure~\ref{fig:performances}. It further allows us to correctly disentangle covariate effects from the precipitation trigger (encoded into the spatial effect), therefore avoiding confusion when the precipitation trigger is not included into the model.
The spatial effect modeling is particularly relevant; a step forward for a better risk management could be made by keeping the estimated covariate coefficients fixed while varying the latent spatial effect by simulating different predictive precipitation scenarios.

Ultimately, we speculate that the estimated latent spatial effect, recall Figure~\ref{fig:SpatialEffect}, provides important information on the spatial distribution of the trigger. For the specific case of the 2009 Messina disaster, researchers have tried to reconstruct the evolution of the storm \citep[e.g.,][]{aronica2012flash}. However, due to the coarse raingauge network in the area, no reliable patterns could have been revealed at a small scale. On the other hand, the latent spatial effect appears to depict a clear spatial trend oriented in the NW-SE direction and centered at the epicentral sector. This information could be used to infer the spatio-temporal evolution of small convective storms in cases where poor weather station coverages represent a physical limitation for rainfall data availability.
Our perspective is that in a small region such as the present study area, the effect of the trigger tends to dominate the slope response compared to the other causative factors. Therefore, we have used the landslide scenario to extract the latent spatial effect, which in turn can provide useful information on the trigger itself.

\section{Conclusion} 
\label{sec:conclusion}
We propose a point process-based modeling framework for debris flow susceptibility with latent random effects embedded into a statistical hierarchical model. Three main novelties arise when contrasting our approach to state-of-the-art approaches currently used in the geomorphological literature. 

The first novelty consists of the prediction of landslide intensity in addition to the more common susceptibility. This provides supplementary information for master planners since we do not only predict where potential landslides could occur under similar climatic stresses, but also how many mass movements might actually take place.

The second novelty concerns the latent spatial effect. Its inclusion in the model strongly increases the  overall prediction performances from excellent to outstanding. We envision future applications of latent spatial effect constructions both to gain better insight into the space-time evolution of the trigger and to simulate alternative predictive landslide  scenarios. In fact, the latent spatial effect model can be numerically simulated with many preferential directions offering supplementary information on how the prediction could change as the trigger pattern changes.

Our third novelty has less practical impact with respect to the previous two, but we believe it to be very important from an interpretative standpoint. It concerns the way covariates are treated in the model. Classical modeling uses continuous covariate in a linear way or discretizes them to a small number of independent categories.  We here investigate differences between linear and nonlinear representations by expressing the latter as a linear effect plus a residual nonlinearity, and we impose prior dependence between adjacent categorical classes. This procedure ensures a relatively smooth regression curve even in cases when some single classes contain only very few data. Moreover, nonlinearity provides a stronger support for interpretative purposes compared to the common linear counterparts, whose use is mainly motivated by computational and modeling convenience. 

Ultimately, we would like to point out the strength of combining pixel- and slope-unit-based models into a hierarchical model structure. For complex methods such as the log-Gaussian Cox point-process models fitted with INLA, one can take advantage from the finely resoluted grid structure, whereas the computationally demanding estimation of the latent spatial effect is done at the slope-unit scale. In summary, this procedure optimizes the structures of the data and model  with respect to the computational costs while maintaining a correct geomorphological and spatial hierarchy.

\section*{Acknowledgement}
Part of the satellite images used to generate the landslide inventory were obtained thanks to the European Space Agency Project (ID: 14151) titled: A remote sensing based approach for storm triggered debris flow hazard modelling: application in Mediterranean and tropical Pacific areas. Principal Investigator: Dr. Luigi Lombardo.

\appendix
\section{Inference through Integrated Nested Laplace Approximation}\label{sec:INLA}
The presence of random effects and the use of Gaussian prior distributions on fixed effect coefficients $\beta_j$ and latent effects $W_{\bm z_k}$ in the latent Gaussian field \eqref{eq:gauss} makes it impossible to directly calculate closed-form expressions of the likelihood or the posterior estimates. We here use the approach of Integrated Nested Laplace Approximation (INLA), implemented in the \texttt{R-INLA} package, for ``integrating out" the latent random components. This approach has already been successfully applied  to regression modeling for Poisson responses and point processes \citep{Illian.al.2012,Taylor.Diggle.2014,Gabriel.al.2016,Opitz.2017}, and the related theory \citep{Rue.Martino.Chopin.2009,Rue.al.2016} confirms that we get accurate approximations of the posterior estimations and predictions in our context of  Poisson regression, in particular when compared to Markov chain Monte Carlo inference which may be hampered by bad mixing of Markov chains for models with complex latent Gaussian structure such as ours. 

While using INLA does not impose any restrictions on the Gaussian variables  in the additive structure of $\log \{\Lambda(s)\}$ that jointly determine a high-dimensional multivariate Gaussian random vector, the number of hyperparameters to be estimated (such as the precision parameters $\tau_k$, $k=1,\ldots, K$, in the spatial effect and the random walk effects) should be kept to a minimum since INLA conducts repeated numerical integrations (the word ``Integrated" in INLA) with respect to the hyperparameter vector. Giving a full presentation of INLA would be far beyond the scope of this paper, but we shortly describe the principal issues that it solves. 
Denote by $\bm x=(x_1,\ldots,x_{n_{\mathrm{grid}}})$  the  latent (Gaussian) log-intensity for the fine-grid pixels such that $\lambda_i=\exp(x_i)$, by $\bm n=(n_1,n_2,\ldots,n_{n_{\mathrm{grid}}})^T$ the observed landslide counts, and by $\bm{\tau} =(\tau_1,\tau_2,\ldots)^T$ the vector of precision hyperparameters. The calculation of univariate posterior densities of $x_i$, $\lambda_i$ and hyperparameters $\tau_k$ is the principal inference goal, but it is hampered by the high-dimensional numerical integration with respect to the latent Gaussian vector $\bm x$: 
\begin{align}
\pi(\tau_k \mid \bm n) &= \int \pi(\bm x, \bm \tau \mid \bm n)\,
\mathrm{d}\bm x\,  \mathrm{d}\bm\tau_{-k}, \label{eq:posttau}\\
\pi(x_i\mid \bm n) &= \int\int \pi(\bm x, \bm\tau\mid \bm n) \mathrm{d}\bm x_{-i}\, \mathrm{d}\bm \tau =  \int \pi(x_i \mid \bm \tau, \bm y)
\pi(\bm \tau\mid \bm n)\,\mathrm{d}\bm\tau, \quad i=1,\ldots,n_{\mathrm{grid}} \label{eq:postlambda}.
\end{align}
While astutely designed numerical integration schemes allow integration with respect to all hyperparameters ($\mathrm{d}\bm\tau$) or all  except for the $k$th one ($ \mathrm{d}\bm\tau_{-k}$), INLA uses Laplace approximation techniques \citep[essentially a Gaussian density approximation of the integrand function, see ][]{Tierney.Kadane.1986} to integrate over the Gaussian components ($\mathrm{d}\bm x$) in the first step, and then in a nested way  in the second step ($\mathrm{d}\bm x_{-i}$, the word ``Nested" in INLA).  

For hyperparameters to be estimated, we must give a nondegenerate prior distribution, while the others must be fixed to a deterministic value (i.e., technically we use a Dirac prior distribution). With highly complex models such as ours, prior distributions should give some guidance to stabilize the estimation procedure by concentrating mass relatively close to a simple reference model; we then let the data decide if significant deviations from the reference model are detected. In our latent Gaussian model, it is natural to concentrate the prior disributions for precision parameters on moderately large values, such that prior mass concentrates around $0$ in the latent Gaussian random vector, the value for ``no effect", i.e., for non-significant fixed covariate effects, for small random walk innovations, and for moderately strong spatial dependence allowing us to borrow strength across neighboring slope units.  	

\baselineskip=10pt

\bibliographystyle{CUP}
\bibliography{landslides}
\end{document}